\documentclass[useAMS,usenatbib,usegraphicx,referee]{mn2e}

\usepackage{amsmath, amssymb, graphics}

\newcommand{\mathsym}[1]{{}}

\title[Magnetic Field Strength in AGN Jets]
{Magnetic Field Strength and Spectral Distribution of Six Active Galactic Nuclei Jets}

\author[O'Sullivan \& Gabuzda]{S. P. O'Sullivan$^{1}$ \& D. C. Gabuzda$^{1}$ \\
$^{1}$Physics Department, University College Cork, Cork, Ireland }

\begin{document}

\date{Released 2008 Xxxxx XX}
\pagerange{\pageref{firstpage}--\pageref{lastpage}} \pubyear{2008}
\maketitle
\label{firstpage}
\begin{abstract}
We use observations of six ``blazars'' with the Very Long Baseline Array (VLBA), at eight frequencies (4.6, 5.1, 7.9, 8.9, 12.9, 15.4, 22.2, 43.1 GHz), to investigate the frequency-dependent position of their VLBI cores (``core-shift'') and their overall jet spectral distribution. By cross-correlating the optically thin jet emission, we are able to accurately align the multi-frequency images of three of the jets (1418+546, 2007+777, 2200+420), whose core-shifts and spectra we find consistent with the equipartition regime of the Blandford \& K\"onigl conical jet model, where the position of the radio core from the base of the jet follows $r_{core}(\nu)\propto \nu^{-1}$. For the jet of 0954+658, we align the higher frequency images using our lower frequency measurements assuming equipartition in the radio core from 4.6--43 GHz. The jet emission of the other two sources in our sample (1156+295, 1749+096) is too sparse for our alignment technique to work. Using our measured core-shifts, we calculate equipartition magnetic field strengths of the order of 10's to 100's of mG in the radio cores of these four AGN from 4.6--43 GHz. Extrapolating our results back to the accretion disk and black hole jet-launching distances, we find magnetic field strengths consistent with those expected from theoretical models of magnetically powered jets.
\end{abstract}

\newpage
\section{Introduction}
Highly collimated jets of relativistic plasma are considered to be generated and propelled outwards from the central regions of Active Galactic Nuclei (AGN) by magnetic forces surrounding the supermassive black hole \citep[see][and references therein]{meierjapan}. High resolution VLBI observations can detect the synchrotron emission from these jets in the radio regime at distances of the order of $10^3-10^7$ $r_g$, where $r_g=GM_{BH}/c^2$, corresponding to the sub-parsec to parsec scales of the jet \citep{lobanovevn}. 
Current theoretical models \citep[][hereafter K07]{vlahakiskonigl2004, komissarov2007} invoke rotation of the black hole magnetosphere and/or the inner accretion disk to create a stiff helical magnetic field that expels, accelerates and helps collimate the jet flow; an external medium is still required to confine the jet at its boundary, with a possible solution being a slower disk-driven wind outflow \cite[e.g.,][]{bogovalovtsinganos2005}. 
The models of \cite{blandfordrees1974} and \cite{dalymarscher1988} show how jets can be accelerated hydrodynamically, however, \cite{konigl2007} proposes that extended acceleration regions are distinguishing characteristics of magnetohydrodynamic (MHD) dominated flows over purely hydrodynamical ones, and uses the model of \cite{vlahakiskonigl2004} to explain the parsec-scale acceleration of component C7 of 3C345 \citep{unwin3c345}. 

In their conical jet model, \cite{blandfordkonigl1979} [hereafter BK79] proposed that ``the unresolved core be identified with the innermost, optically thick region of the approaching jet'' with the base of the jet corresponding to the vertex of the cone. The radio cores of AGN generally have flat or inverted ($\alpha>0$) spectra, where $S\propto\nu^{\alpha}$. The optical depth due to synchrotron self-absorption (SSA) is described by $\tau(r)\propto r^{(1.5-\alpha)m+n-1}\nu^{2.5-\alpha}$ \cite[e.g.,][]{rybickilightman1979}, where $m$ and $n$ describe the power-law fall-off in the magnetic field strength ($B=B_{1\rm{pc}}r^{-m}$) and particle number density ($N=N_{1\rm{pc}}r^{-n}$), respectively, with distance from the central engine. Since VLBI resolution is usually not sufficient to completely resolve the true optically thick radio core, the core region contains emission from around the $\tau=1$ surface at that frequency, as well as some contribution from the optically thin inner jet. 

Furthermore, due to the frequency dependence of the $\tau=1$ surface, the position of the radio core follows $r\propto\nu^{-1/k_{r}}$ \citep{konigl1981}, where $r$ is the distance from the central engine and $k_r=((3-2\alpha)m+2n-2)/(5-2\alpha)$. For equipartition between the jet particle and magnetic field energy densities, the quantity $k_r=1$, with the choice of $m=1$ and $n=2$ being physically reasonable \citep[e.g.,][hereafter L98]{konigl1981, huttermufson1986, lobanov1998}, making $k_r$ independent of the spectral index. Therefore, in equipartition, the ``core-shift'' ($\Delta r=r_{\nu_{1}}-r_{\nu_{2}}$) between two frequencies ($\nu_{2}>\nu_{1}$) is directly proportional to $(\nu_{2}-\nu_{1})/(\nu_{2} \nu_{1})$.

Since absolute positional information is lost during calibration of VLBI images, they are centered at the peak intensity of the map, which does not necessarily correspond to the furthest upstream component referred to as the VLBI core. For accurate combination of multi-frequency VLBI maps, we must properly align our images by correcting for the frequency-dependent position of the self-absorbed radio core. 
Phase-reference observations \citep{marcaideshapiro1984} can be used to obtain the absolute position of the radio core by comparing the position of the target source with respect to a reference source, which should ideally be a point source, but this is a very calibration-intensive and weather dependent approach and gets much more difficult with observations at more than one frequency.
The frequency dependent core-shift has generally been obtained by aligning model-fitted optically thin jet components at different frequencies (``self-referencing'') \citep[e.g.,][]{lobanov1998, kovalev2008}, but this method is unsuitable for smooth or weak jets that do not have a distinct component across a wide range of frequencies on which to align the images. 
An alternative method of aligning VLBI images is to cross-correlate the optically thin emission from the jet at multiple pairs of frequencies from regions where we do not expect large variations in the spectral index. \cite{walker} used such a cross-correlation technique to successfully align his multi-frequency images of 3C84.
We also use a cross-correlation technique for aligning our images, which involves comparing the optically thin jet emission at two frequencies with the highest correlation-coefficient occurring when the two images are best aligned; see \cite{crokegabuzda2008} for a detailed description of the method. 
From this method, we have been able to find reliable measurements for the frequency dependent core-shift of several sources across a wide range of frequencies \citep[see][for comparisons with both phase referencing and ``self-referencing'' methods]{osullivangabuzda9thevn}.

From long-term monitoring of blazar jets, \cite{marschervsop2009} lists evidence supporting the idea that in some cases the cores correspond to a stationary feature such as a conical shock \citep{cawthorne2006}. He states that the core would correspond to either the $\tau=1$ surface or the first stationary hotspot downstream of that surface, whichever is more intense at a particular epoch. Our core-shift measurements after the alignment of the multi-frequency images of 2200+420, 2007+777, 1418+546 and 0954+658 are consistent with the assumption of the core corresponding to the $\tau=1$ surface and not some stationary feature. It's also worthwhile inspecting the core polarization because this can sometimes give information about scales smaller than the standard resolution at a particular frequency. From the polarization images of these sources presented in \cite{osullivangabuzda2009}, we do not see any clear evidence of distinct polarized features upstream of the total intensity core which would support the model of a stationary core feature through which ejected plasma blobs propagated. 

In Section 2 we describe our observations and how we use our core-shift measurements to calculate magnetic field strengths. Our results for each source are presented in Section 3 and we discuss some of their interesting features in Section 4. Our conclusions are listed in Section 5. 


\section{Observations and Method}
Six ``blazars'' were observed on 2006 July 2 with the American Very Long Baseline Array (VLBA) at eight frequencies (4.6, 5.1, 7.9, 8.9, 12.9, 15.4, 22.2, 43.1 GHz) quasi-simultaneously \citep[see][for details]{osullivangabuzda2009}. Here we present spectral index images for these six blazars along with estimates of the magnetic field strengths in the radio core from 4.6--43 GHz for four sources in which we were able to measure the ``core-shift''. We follow the method described in L98 for calculating several important jet parameters from the frequency-dependent shift of the position of the VLBI core due to SSA. Table \ref{source2200} lists the observed sources along with the model parameters used in our calculations.

\begin{table}
 \caption{Model Parameters for all sources}
 \label{source2200}
 \begin{tabular}{@{}cccccccc}
  \hline
  Source      &   $z$     & $D_L$ & $\alpha_j$    & $\phi$      & $\Gamma$   & $\theta$    &  $\delta$ \\
                   &              & (Mpc)  &                     & $(^\circ)$ &                      & $(^\circ)$   &               \\
  \hline
  0954+658 &  0.368   & 1958     & $-0.6$       &    2.2      &        6.3$^{\ast}$       &     9.3          &  6.2$^{\dagger}$  \\
  1156+295 &  0.729   & 4481     &   -       &      -        &        25.1$^{\dagger}$     &    2.0$^{\dagger}$               &  28.5$^{\dagger}$  \\
  1418+546 &  0.152   &   716     & $-0.5$       &    1.9      &         4.5$^{\star}$      &    11.2        &  5.1$^{\dagger}$   \\
  1749+096 &  0.320   & 1662     &   -       &      -        &        7.5$^{\dagger}$       &    3.8$^{\dagger}$             &  12.0$^{\dagger}$   \\
  2007+777 &  0.342   & 1797     & $-0.5$       &    1.0      &         8.0      &     7.2         &   7.9$^{\dagger}$  \\
  2200+420 &  0.069   &   307     & $-0.5$       &    1.9$^{\ddagger}$      &        7.0$^{\ddagger}$       &    7.7$^{\ddagger}$         &   7.2$^{\ddagger}$  \\
  \hline
 \end{tabular}
 \\{Cosmology: H$_0 = 71$ km s$^{-1}$ Mpc$^{-1}$, $\Omega_{\Lambda}=0.73$ and $\Omega_{M}=0.27$. Col.~(1): IAU source name, Col.~(2): redshift, Col.~(3): luminosity distance, Col.~(4): jet spectral index, Col.~(5): jet half-opening angle, Col.~(6): bulk Lorentz factor, Col.~(7): jet viewing angle, Col.~(8): Doppler factor. References: $^{\ast}$\cite{gabuzda1994}, $^\star$\cite{pushkarev1999}, $^\dagger$\cite{hovatta2009}, $^\ddagger$\cite{jorstad2005}}
\end{table}

Once our images are properly aligned, we generate spectral index ($\alpha$) maps between pairs of frequencies in {\sevensize AIPS} using the task {\sevensize COMB}, with $\alpha=(\log S_{\nu_2} - \log S_{\nu_1})/(\log \nu_2 - \log \nu_1)$ where $S_{\nu}$ is the measured flux at frequency $\nu$. This approximates the spectrum as a straight line which should be a good approximation as long as there is not a spectral turnover between the two frequencies. Spectral turnovers of components are often observed in VLBI jets \citep{lobanovspectra1998} but useful information can still be extracted by generating maps from relatively closely spaced frequency pairs. In order to analyse the spectrum across our wide frequency range, we constructed spectral index maps from 4.6--7.9, 8.9--12.9, 15.4--22 and 22--43 GHz for each source. The images used to create each map have the same image size, pixel size and are convolved with the same beam, corresponding to the lower frequency of the pair. We also take a slice along the jet of each spectral index image for inspection.

Since the shift measurements from the cross-correlation program are found with respect to the Stokes {\emph I} peak intensities, we use them in conjunction with the model-fitted core positions from {\sevensize DIFMAP} \citep{shepherd1997} to obtain the distance between the cores at different frequencies (``core-shift''). The task {\sevensize UVRANGE} in {\sevensize DIFMAP} was used to provide {\emph uv}--ranges for the higher frequencies matching that of the lower frequencies when model-fitting. Due to the fall-off in the optically thin part of the synchrotron spectrum with increasing frequency and the large differences in resolution, aligning images across wide spacings in frequency can prove very difficult. However, with our excellent frequency coverage from 4.6--43 GHz, we are able to obtain many measurements from different frequency combinations which can be compared with each other to check for consistency.

To calculate the magnetic field strength with distance from the base of the jet, we combine the analysis of L98 and \cite{hirotani2005}.
The distance of the radio core from the base of the jet at the observed frequency $\nu$ in GHz, is given by
\begin{equation}
\label{rcore}
r_{\rm core}(\nu)=\left(\frac{\Omega_{r\nu}}{\sin \theta} \right)  \nu^{-1/k_r}
\end{equation}
in parsecs, where $\theta$ is the jet viewing angle and $\Omega_{r\nu}$ is the core-position offset measure, defined in L98 as
\begin{equation}
\label{omega}
\Omega_{r\nu}=4.85\times10^{-9} \frac{\Delta r_{\rm mas} D_L}{(1+z)^2} \left( \frac{\nu_1^{1/k_r}\nu_2^{1/k_r}}{\nu_2^{1/k_r}-\nu_1^{1/k_r}} \right)
\end{equation}
in units of ${\rm pc}\cdot{\rm GHz}$, where $D_L$ is the luminosity distance in parsecs and $\Delta r_{\rm mas}$ is the measured core-shift in milliarcseconds.

We use the synchrotron self-absorption derivation for a conical jet by \cite{hirotani2005} and the equipartition condition, in cgs units,
\begin{equation}
\label{equipartition}
N\gamma_{\rm min}m_e c^2=K \frac{B^2}{8\pi}
\end{equation}
\begin{equation}
  K = \frac{2\alpha+1}{2\alpha}
         \frac{(\gamma_{\rm max}/\gamma_{\rm min})^{2\alpha  }-1}
              {(\gamma_{\rm max}/\gamma_{\rm min})^{2\alpha+1}-1}.
\end{equation}
where $\gamma_{\rm min}$ and $\gamma_{\rm max}$ are the low and high cut-offs of the electron energy distribution.  
Evaluating eqn.~43 from \cite{hirotani2005} for the simplified case $k_r=1$ and $\alpha=-0.5$, 
we get the equipartition magnetic field in Gauss at 1~pc
\begin{equation}
\label{B1}
B_1\simeq0.025 \left( \frac{\Omega_{r\nu}^{3} (1+z)^2}{ \delta^2 \phi \sin^2\theta} \right)^{1/4}
\end{equation}
where $\delta$ is the Doppler factor, $\theta$ is the viewing angle and $\phi$ is the half-opening angle of the jet. The magnetic field strength in the core at the observed frequency $\nu$ can then be found from
\begin{equation}
\label{Bcore}
B_{\rm core}(\nu)=B_1 r_{\rm core}^{-1}
\end{equation}

For spectral indices other than $-0.5$, the magnetic field strength depends on the exact value of $\gamma_{\rm min}$ according to $B_1\propto\gamma_{\rm min}^{(2+4\alpha)/(7-2\alpha)}$. The dependence on $K$ is much weaker since $\gamma_{\rm max} \gg \gamma_{\rm min}$. Equation~\ref{B1} differs from the equipartition magnetic field at 1 pc derived in L98, eqn.~10, where our calculated magnetic field strengths are a factor of $\sim3$ times smaller (e.g., compare our results for 2200+420 here with those in \cite{osullivangabuzda9thevn}) due to the more accurate equipartition condition used in \cite{hirotani2005}. Note that in all cases, where errors are presented, we have only considered the errors in the core-shift measurements.

\section{Magnetic Field Strength and Spectral Index Results}
\subsection{2200+420 (BL Lac)}
\begin{figure*}
\includegraphics[width=180mm]{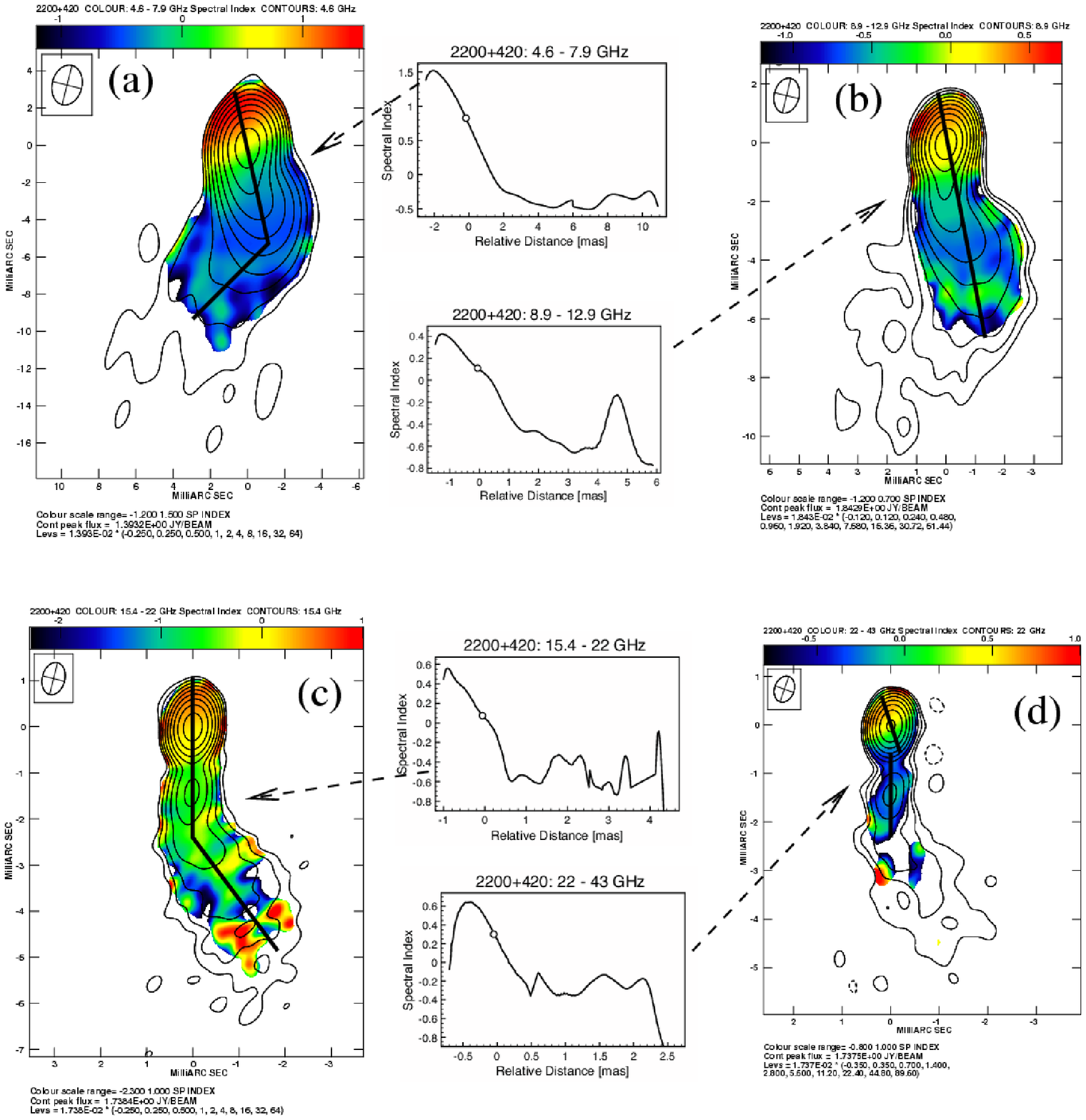}
 \caption{Spectral index maps for 2200+420 constructed from frequency pairs, with plots of slices along the jet region indicated by the solid line.
 (a) 4.6--7.9 GHz. Spectral index scale range: $-1.2$ to 1.5.
 Clip levels of {\emph I} maps: 3.5 mJy (4.6 GHz), 3.8 mJy (7.9 GHz).
 Beam FWHM and Position Angle: 2.17$\times$1.60 mas, $-14^{\circ}$ (4.6 GHz).
 Contours: 4.6 GHz {\emph I}, starting at 3.48 mJy and increasing by factors of 2.
 (b) 8.9--12.9 GHz. Spectral index scale range: $-1.2$ to 0.7.
 Clip levels of {\emph I} maps: 2.2 mJy (8.9 GHz), 6.8 mJy (12.9 GHz).
 Beam FWHM and Position Angle: 1.21$\times$0.84 mas, $-14^{\circ}$ (8.9 GHz).
 Contours: 8.9 GHz {\emph I}, starting at 2.21 mJy and increasing by factors of 2.
 (c) 15.4--22 GHz. Spectral index scale range: $-2.3$ to 1.0.
 Clip levels of {\emph I} maps: 5.2 mJy (15.4 GHz), 5.9 mJy (22 GHz).
 Beam FWHM and Position Angle: 0.71$\times$0.49 mas, $-14^{\circ}$ (15.4 GHz).
 Contours: 15.4 GHz {\emph I}, starting at 4.35 mJy and increasing by factors of 2.
 (d) 22--43 GHz. Spectral index scale range: $-0.8$ to 1.0.
 Clip levels of {\emph I} maps: 6.9 mJy (22 GHz), 13.3 mJy (43 GHz).
 Beam FWHM and Position Angle: 0.55$\times$0.40 mas, $-19^{\circ}$ (22 GHz).
 Contours: 22 GHz {\emph I}, starting at 6.08 mJy and increasing by factors of 2.
 The arrows connect the plots of the spectral index slices to their corresponding maps. The hollow circle on the slice plot indicates the core position at the lower frequency of the pair.}
 \label{2200spix}
\end{figure*}

Fig.~\ref{2200spix} shows the spectral index distribution for 2200+420 from 4.6--43 GHz. The core region displays a strongly inversted spectrum at low frequencies (Fig.~\ref{2200spix}a) before becoming less inverted at higher frequencies (Fig.~\ref{2200spix}b, c, d), consistent with SSA of emission from the unresolved part of the jet. Further downstream, the jet emission becomes optically thin, with the majority of slices consistent with a spectral index of $\sim-0.5$. The direction of the slice along the core region in the 22--43 GHz map (Fig.~\ref{2200spix}d) follows the 43-GHz inner jet direction. 
There is an interesting feature in the 8.9--12.9 GHz map (Fig.~\ref{2200spix}b), at $\sim$5 mas from the core, where the spectrum flattens considerably. This feature is also present in the 4.6--7.9 GHz and 15.4--22GHz maps, but is less pronounced, possibly due to the low resolution at 4.6 GHz and the lack of sensitivity to the extended jet emission at the higher frequencies. The presence of this feature might be due to some particle re-acceleration in the jet or interaction between the jet and the surrounding medium, since this is close to where the jet changes to a south-easterly direction in the plane of the sky.

The measured core-shifts for 2200+420 are listed in Table~\ref{shifts2200} along with the angle representing the direction of the shift measured from north through east. The average direction of the shift is $14\pm10^{\circ}$, where the error corresponds to the standard deviation, and this core-shift direction is consistent with the inner jet direction at 43 GHz of $\sim-163^{\circ}$. Table~\ref{43ref2200} lists the size of the shift as a fraction of the beamsize. Even though the shifts are only of the order of 5-20\% of the corresponding beams, they still have quite a dramatic effect on the spectral index in the core if they are not taken into account.

\begin{figure}
\includegraphics[width=90mm]{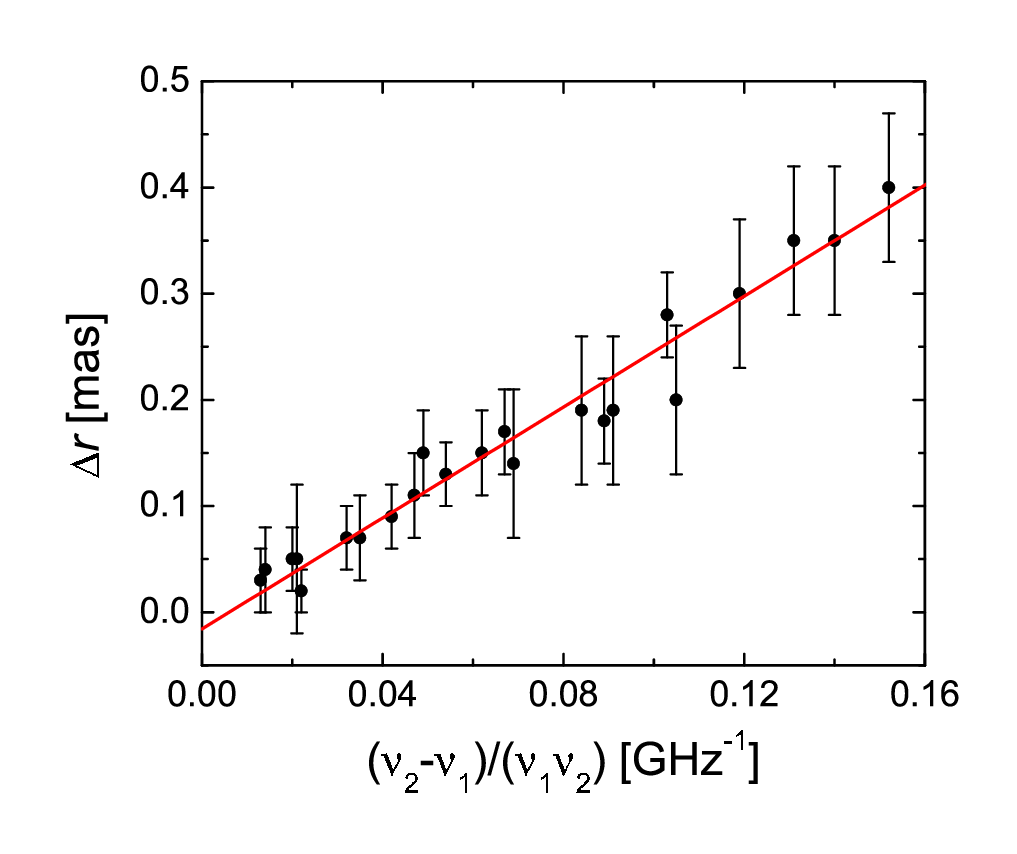}
 \caption{Straight-line fit to the all the core-shift data for 2200+420, assuming equipartition. The y-intercept value is equal to $-0.01\pm0.01$ mas. A straight-line fit through the origin is expected for $k_r=1$.}
 \label{all2200}
\end{figure}

\begin{figure}
\includegraphics[width=90mm]{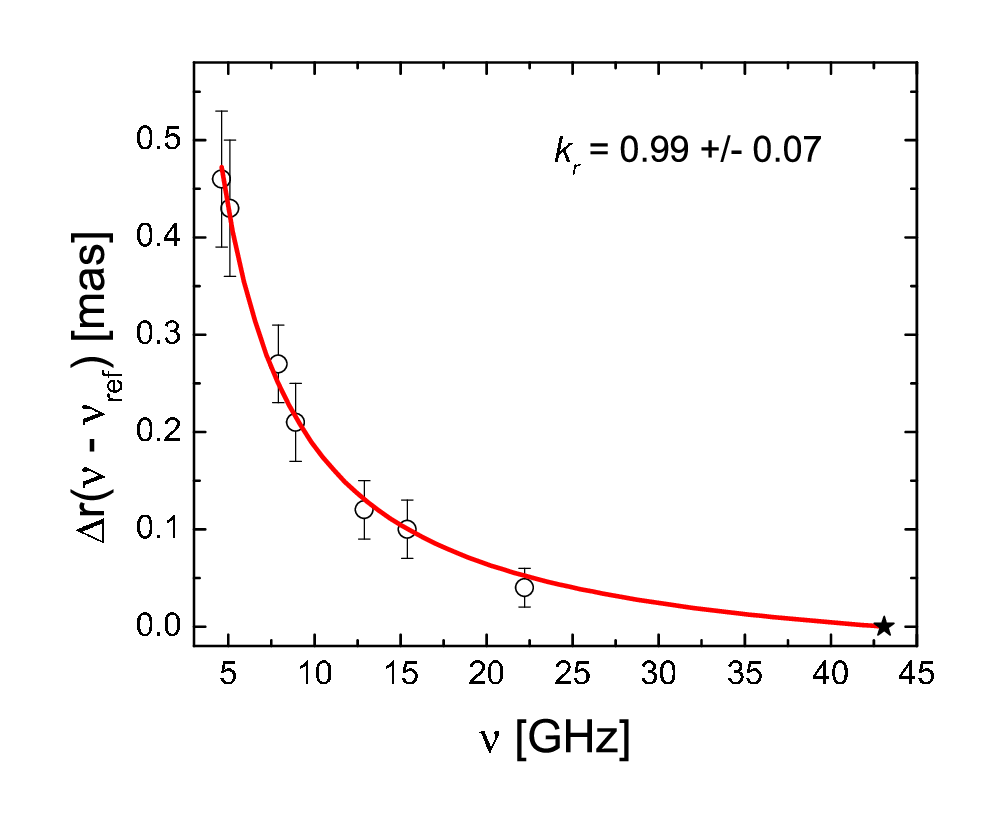}
 \caption{Plot of core-shift measurements versus frequency for 2200+420 (listed in Table \ref{43ref2200}), using 43 GHz as the reference frequency. Best-fit line: $\Delta r=A(\nu^{-1/k_r}-43.1^{-1/k_r})$ with parameters $A=2.5\pm0.3$ and $k_r=0.99\pm0.07$}
 \label{kr2200}
\end{figure}

Figure~\ref{all2200} shows each shift measurement, $\Delta r$, plotted against the expected frequency dependence for $k_r=1$. The y-intercept from a straight line fit to the data is $-0.01\pm0.01$ mas. From the slope of the line we determine a characteristic core position offset measure ($\Omega_{r\nu}$) of $3.4\pm0.3$ ${\rm pc}\cdot{\rm GHz}$ for 2200+420.
Our results are also consistent with a previous core-shift estimate from \cite{mutel1990} who found a shift of 0.3 mas between 5.0 and 10.6 GHz, corresponding to $\Omega_{r\nu}=3.7$ ${\rm pc}\cdot{\rm GHz}$ (no error was presented with their result).
Since the position of the core is expected to change with frequency according to $r_{\rm core}(\nu)\propto \nu^{-1/k_{r}}$, observations at three or more frequencies enable determination of the parameter $k_r$. Our dense frequency sampling allows us to provide strong constraints on the value of $k_r$ across our entire frequency range. Using 43 GHz as the reference frequency and averaging the various possible combinations of shift measurements (Table~\ref{43ref2200}), we find our values to be consistent with a constant $k_r$ of $0.99\pm0.07$ from 4.6--43GHz (Fig.~\ref{kr2200}). This indicates that the physical conditions in the compact jet region in 2200+420 do not change significantly between 4.6 and 43 GHz. A value of $k_r=1$ indicates that the jet particle and magnetic field energy densities are in equipartition, and our derived value of $k_r$ is consistent with this assumption. 

\begin{table}
 \caption{All measured core-shifts for 2200+420, plotted in Fig.~\ref{all2200}}
 \label{shifts2200}
 \begin{tabular}{@{}cccccccc}
  \hline
  $\nu_1$ & $\nu_2$ & $\Delta r$ &   Angle          \\
    (GHz)  &  (GHz)    &   (mas)     &   ($^{\circ}$)   \\
  \hline
  4.6 & 5.1 & $0.06\pm0.07$ & $-1.0\pm0.1$  \\
  4.6 & 7.9 & $0.19\pm0.07$ & $15.2\pm0.3$  \\ 
  4.6 & 8.9 & $0.20\pm0.07$ & $15.5\pm0.3$  \\
  4.6 & 12.9 & $0.35\pm0.07$ & $19.4\pm0.4$  \\
  4.6 & 15.4 & $0.40\pm0.07$ & $16.3\pm0.3$  \\
  5.1 & 7.9 & $0.14\pm0.07$ & $22.0\pm0.6$  \\
  5.1 & 8.9 & $0.19\pm0.07$ & $16.4\pm0.3$  \\
  5.1 & 12.9 & $0.30\pm0.07$ & $23.3\pm0.6$  \\
  5.1 & 15.4 & $0.35\pm0.07$ & $19.1\pm0.4$  \\
  7.9 & 8.9 & $0.04\pm0.04$ & $5.6\pm0.1$ \\
  7.9 & 12.9 & $0.15\pm0.04$ & $18.8\pm0.3$  \\
  7.9 & 15.4 & $0.18\pm0.04$ & $14.0\pm0.1$  \\
  7.9 & 22.2 & $-$ & $-$  \\
  7.9 & 43.1 & $0.28\pm0.04$ & $6.4\pm0.1$  \\
  8.9 & 12.9 & $0.07\pm0.04$ & $11.0\pm0.1$  \\
  8.9 & 15.4 & $0.11\pm0.04$ & $20.9\pm0.3$  \\
  8.9 & 22.2 & $0.17\pm0.04$ & $13.1\pm0.1$  \\
  8.9 & 43.1 & $0.18\pm0.04$ & $-0.6\pm0.1$  \\
  12.9 & 15.4 & $0.03\pm0.03$ & $-9.1\pm0.1$  \\
  12.9 & 22.2 & $0.07\pm0.03$ & $12.7\pm0.1$  \\
  12.9 & 43.1 & $0.13\pm0.03$ & $2.3\pm0.1$  \\
  15.4 & 22.2 & $0.05\pm0.03$ & $23.5\pm0.3$  \\
  15.4 & 43.1 & $0.09\pm0.03$ & $19.9\pm0.2$  \\
  22.2 & 43.1 & $0.03\pm0.02$ & $33.2\pm0.4$  \\ 
  \hline
 \end{tabular}
\\Col.~(1) \& (2): observed frequency in GHz, Col.~(3): magnitude of the core-shift in milliarcseconds, Col.~(4): angle of core-shift in degrees. Average angle: $14\pm10^{\circ}$
\end{table}

\begin{table}
 \caption{Averaged core-shifts using 43 GHz as the reference frequency for 2200+420, plotted in Fig.~\ref{kr2200}}
 \label{43ref2200}
 \begin{tabular}{@{}cccc}
  \hline
  $\nu$ & $\Delta r$ &  $\Delta r_{projected}$ & Fraction of beam \\
   (GHz) &   (mas)    &       (pc)       &   (\%)     \\
  \hline
  4.6 & $0.46\pm0.07$ & $0.60\pm0.09$ & 17\\
  5.1 & $0.43\pm0.07$  & $0.56\pm0.09$ & 17\\
  7.9 & $0.27\pm0.04$  & $0.35\pm0.05$ & 16\\
  8.9 & $0.21\pm0.04$  & $0.27\pm0.05$ & 14\\
  12.9 & $0.12\pm0.03$  & $0.16\pm0.04$ & 11\\
  15.4 & $0.10\pm0.03$  & $0.13\pm0.04$ & 10\\
  22.2 & $0.04\pm0.02$  & $0.05\pm0.03$ & 5\\
  \hline
 \end{tabular}
\\Col.~(1): observed frequency in GHz, Col.~(2): magnitude of the shift from the 43-GHz core position, Col.~(3): projected shift in parsecs (1~mas = 1.3 pc), Col.~(4): core-shift as a fraction of the beam-size at the corresponding frequency, in percent.
\end{table}

Using our core-shift measurements, we can find estimates for the magnetic field strength, distance of the radio core to the base of the jet and the total jet luminosity. To do this we follow the method outlined in L98. Estimates of the jet opening angle, viewing angle, bulk Lorentz factor and Doppler factor are taken from multi-epoch VLBI observations of 2200+420 
by \cite{jorstad2005}, see Table~\ref{source2200}. These estimates are also consistent with more recent observations by \cite{hovatta2009}, who analyse the variability of radio flares from single dish measurements to obtain their jet parameters. We use a jet spectral index ($\alpha_j$) of $-0.5$ for our calculations, consistent with our spectral index maps for this source (Fig.~\ref{2200spix}). 
We use the shift measurements listed in Table~\ref{shifts2200} to calculate the average core-position offset measure ($\Omega_{r\nu}$) from eqn.~\eqref{omega} at each frequency. 
Then for each $\Omega_{r\nu}$, we calculate the de-projected distance of the radio core to the base of the jet, $r_{core}(\nu)$ from eqn.~\eqref{rcore}, the equipartition magnetic field strength at 1 pc, $B_1$ from eqn.~\eqref{B1}, and finally, using eqn.~\eqref{Bcore} the magnetic field strength in the radio core. Our results are listed in Table~\ref{results2200}.

It's important to note the effect that small changes in the parameter $k_r$ have on our calculated values. We get $0.92<k_r<1.06$ from the range of possible values within the error in the fit shown in Fig.~\ref{kr2200}. There is a strong dependence on the value of $r_{core}(\nu)$ with $3.5<r_{core}(43{\rm~GHz})<0.14$ pc, while the magnetic field strength at 1 pc remains relatively insensitive to small variations in $k_r$ with $0.12<B_1<0.14$ pc (using $\Omega_{r\nu}$ at 43~GHz). Therefore, while there is a rather large uncertainty in $r_{core}$ and hence $B_{core}$, we consider the calculated value of $B_1$ to be reliable for the jet parameters listed in Table~\ref{source2200}.

\begin{figure}
\includegraphics[width=100mm]{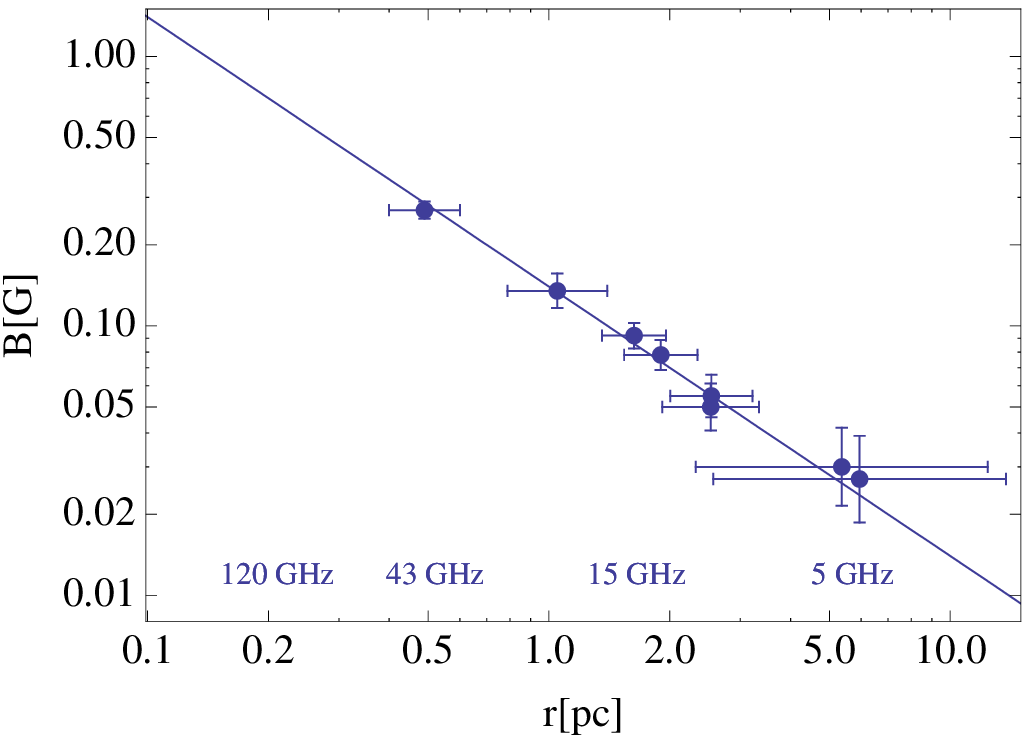}
 \caption{Plot of magnetic field strength ($B$) in units of Gauss versus distance along the jet ($r$) in units of parsecs for the core of 2200+420 at 4.6, 5.1, 7.9, 8.9, 12.9, 15.4, 22.2 \& 43 GHz using $k_r=1$. Also plotted is the best-fit line $B=B_1r^{-1}$ with $B_1=0.139$ G with a formal error of 0.006 G.}
 \label{bvr2200}
\end{figure}

Adopting $k_r=1$, we plot the magnetic field strength in the core as a function of the distance from the base of the jet in Fig.~\ref{bvr2200}. Fitting the line $B(r)=B_1r^{-1}$ to the data, we find $B_1=0.139$ G with a formal error of $0.006$ G. By extrapolation, our results predict field strengths of the order of 1 G in the 1 mm core of 2200+420. This is consistent with results for other jets from \cite{savolainenbologna2008}, where he finds magnetic field strengths of order 1 G in the core of 3C~273 by measuring the synchrotron self-absorption turnover frequency \citep[e.g.,][]{marscher1987} of individual model-fitted components from 5--86 GHz (60--3.5 mm). Importantly, this method does not require the assumption of equipartition to separate the magnetic field strength from the particle number density, but the calculations are very sensitive to the exact values of the turnover frequency and the estimated size of the model-fitted core component at the turnover frequency.

In their simulations of magnetically driven relativistic jets, K07 consider applications of their model to AGN jets and provide a means of estimating the distance from the central engine at which equipartition between the Poynting and matter energy fluxes is established, $R_{\rm eq,BH}\sim2\times10^{16}(M_{\rm BH}/10^8M_{\odot})(0.1/\phi)$ cm, for jets launched from the black hole or from the nuclear accretion disk $R_{\rm eq}(AD)\sim10^{17}(M_{\rm BH}/10^8M_{\odot})(0.1/\phi)$ cm. In their model, the jet flow is magnetically accelerated up to a distance of $\sim10R_{eq}$, at which point it reaches the terminal Lorentz factor; hence, observations that probe this region may detect deviations from the jet model of BK79, which assumes a constant jet speed, and possibly infer whether the observed jet has been launched from the black hole or the accretion disk. 
Using the jet opening angle of $1.9^{\circ}$ and a black hole mass of $1.7\times10^8 M_\odot$ \citep{woourry2002}, the equipartition distance for a black-hole-launched jet in 2200+420 is $\sim0.03$ pc while $R_{eq}\sim0.2$ pc for a jet launched from the accretion disk, which according to our results, would be accessible with VLBI observations at frequencies greater than 120 GHz. This is out of the range of current VLBI but could be probed with future sub-mm VLBI arrays \citep[e.g.,][]{doeleman2009}. However, since the terminal Lorentz factor is not reached until $\sim10R_{eq}$ it's possible that accelerating jet components could be seen in the inner jet at frequencies as low as $\sim15$ GHz. Multi-epoch VLBA observations at 43 GHz from \cite{jorstad2005} have indeed found accelerating jet components for this source which they attribute to a small change in the Lorentz factor along with a substantial change in the direction of the jet flow. The large uncertainty in $r_{core}$ should be kept in mind though, as $r_{core}(43{\rm~GHz})$ may extend as far as 3.5 pc from the base of the jet.

Using the divergence theorem, we can also estimate the initial magnetic field strength at the jet launching distance by assuming the energy extracted per time is equal to the Poynting flux times the area (N. Vlahakis, private communication), which gives $B_0\sim1.5\times10^{-5}L^{1/2}r^{-1}$. For example, with $L\sim10^{46}$ erg s$^{-1}$ and $M_{\rm BH}\sim10^8$ M$_\odot$, we get $B_{\rm 0,AD}\sim10^4$ G at an inner accretion disk distance of $10r_g$ and $B_0(BH)\sim10^5$ G at a black hole jet-launching distance of $r_g$, as quoted in K07.

\begin{figure}
\includegraphics[width=100mm]{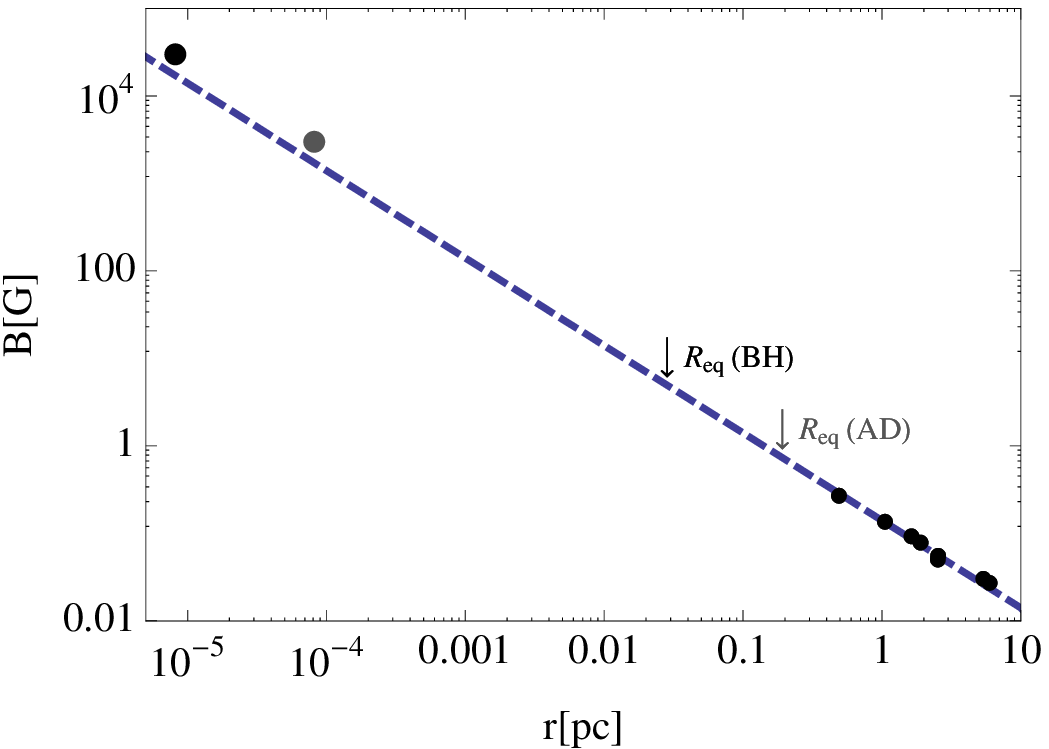}
 \caption{Extended plot of $B$ versus $r$ for 2200+420 to include the extrapolated magnetic field strength ($B\simeq0.14r^{-1}$) at the accretion disk (10$r_g$) and black holes ($r_g$) distances. The estimated equipartition radii from K07 are indicated by the arrows with $R_{\rm eq,AD}\sim0.2$ pc and $R_{\rm eq,BH}\sim0.03$ pc. Also plotted are points at 10$r_g$ (grey) and $r_g$ (black) representing the theoretically expected magnetic field strengths for a magnetically powered jet.}
 \label{bvrext2200}
\end{figure}

Using eqn.~(23) from BK79, we can find an estimate for the total power carried by the relativistic electrons and magnetic field in the jet, $L_{\rm total}\sim2.5\times10^{48}\Gamma^2 \beta \phi^2 B_{1}^{2}$ erg s$^{-1}$, where $\beta$ is the speed of the jet in units of $c$. From our best-fit value of $B_1\sim0.14$ G, we get $L_{\rm total}\sim(2.6\pm0.2)\times10^{45}$ erg s$^{-1}$. Hence, for 2200+420, we expect $B_0(AD)\sim3\times10^3$ G and $B_0(BH)\sim3\times10^4$ G.
In Fig.~\ref{bvrext2200}, we extend our plot of $B$ vs. $r$ to include the calculated values of the initial magnetic field strength at the corresponding jet launching distances.
Extrapolating $0.14r^{-1}$ all the way down to $1r_g$, we find $B\sim1.7\times10^4$ G, in good agreement with the estimated magnetic strengths at the nominal jet launching distances.


\begin{table}
 \caption{Derived physical quantities for 2200+420 ($k_r=1$)}
 \label{results2200}
 \begin{tabular}{@{}cccccccc}
  \hline
  $\nu$ & $B_{core}(\nu)$ & $r_{core}(\nu)$ & $B_{1\rm{pc}}$ &  $\Omega_{r \nu}$\\
   (GHz) &       (G)             &        (pc)            &          (G)         & (${\rm pc} \cdot {\rm GHz}$) \\
  \hline
  4.6    & $0.03\pm0.01$     & $5.9\pm5.0$    & $0.16\pm0.20$    &    -\\
  5.1    & $0.03\pm0.01$     & $5.4\pm4.5$    & $0.16\pm0.20$    &    $3.7\pm3.1$\\
  7.9    & $0.05\pm0.01$     & $2.5\pm0.7$    & $0.13\pm0.05$    &    $2.8\pm0.7$\\
  8.9    & $0.06\pm0.01$     & $2.5\pm0.6$    & $0.14\pm0.05$    &    $3.0\pm0.8$\\
  12.9   & $0.08\pm0.01$     & $1.9\pm0.4$    & $0.15\pm0.05$    &   $3.3\pm0.7$\\
  15.4   & $0.09\pm0.01$     & $1.6\pm0.3$    & $0.15\pm0.04$    &   $3.4\pm0.6$\\
  22.2   & $0.14\pm0.02$     & $1.1\pm0.3$    & $0.14\pm0.06$    &   $3.1\pm0.9$\\
  43.1   & $0.27\pm0.02$     & $0.5\pm0.1$    & $0.13\pm0.03$    &   $2.8\pm0.4$\\
  \hline
 \end{tabular}
\\Col.~(1): observed frequency in GHz, Col.~(2): radio core magnetic field strength in Gauss, Col.~(3): de-projected distance of the radio core to the base of the jet in parsecs, Col.~(4): magnetic field strength at 1 pc, Col.~(5): average core-position offset measure, in units of ${\rm pc} \cdot {\rm GHz}$, for shifts with $\nu_2$ in Table~\ref{shifts2200} equal to $\nu$ in Col.~(1) above.
Note: Physical parameters at 4.6~GHz calculated with 5.1 GHz core-position offset measure.
\end{table}

\subsection{2007+777}
\begin{figure*}
\includegraphics[width=150mm]{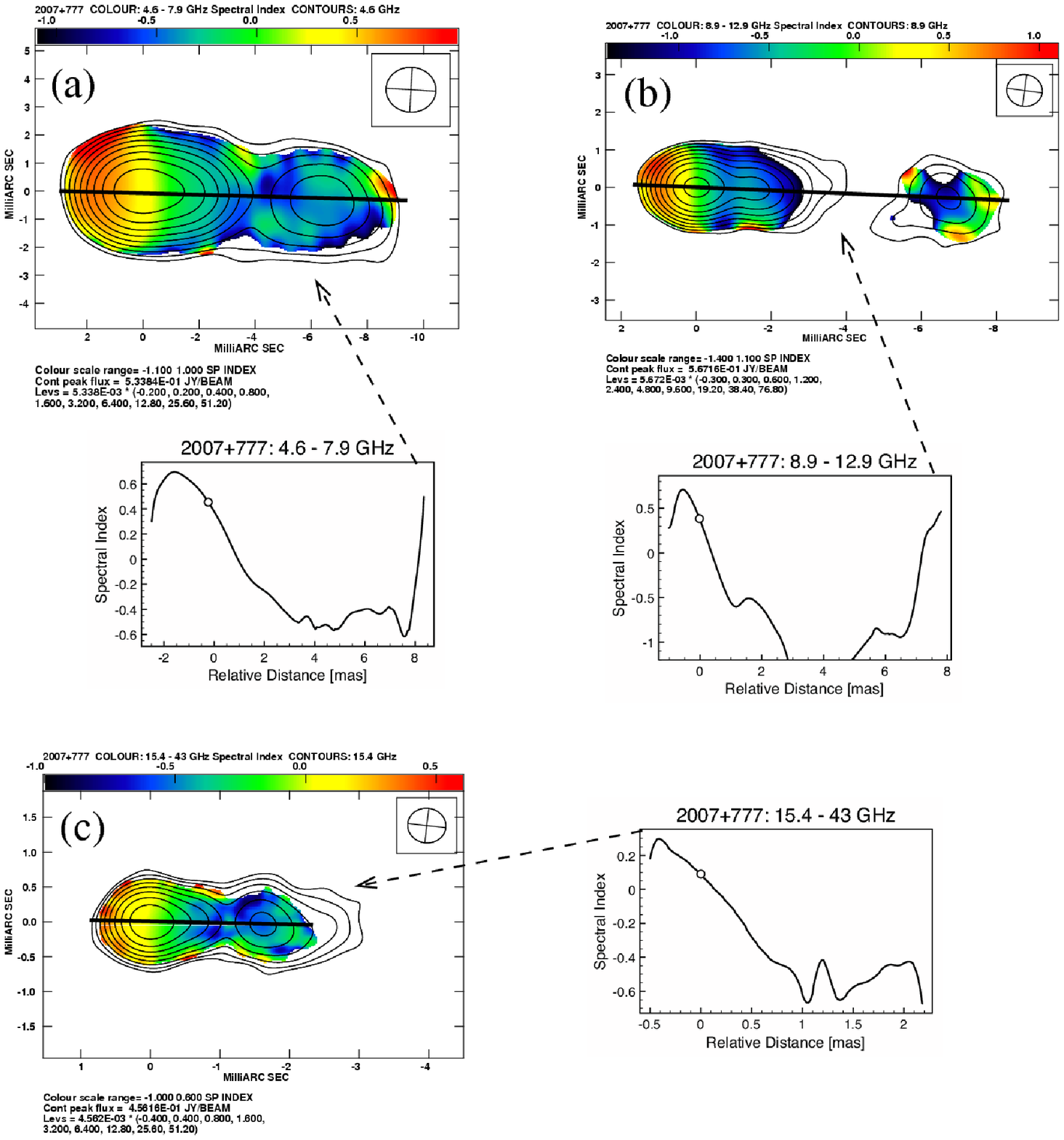}
 \caption{Spectral index maps for 2007+777 constructed from frequency pairs, with plots of slices along the jet region indicated by the solid line.
 (a) 4.6--7.9 GHz. Spectral index scale range: $-1.1$ to 1.0.
 Clip levels of {\emph I} maps: 1.1 mJy (4.6 GHz), 2.0 mJy (7.9 GHz).
 Beam FWHM and Position Angle: 1.78$\times$1.60 mas, $86.5^{\circ}$ (4.6 GHz).
 Contours: 4.6 GHz {\emph I}, starting at 1.07 mJy and increasing by factors of 2.
 (b) 8.9--12.9 GHz. Spectral index scale range: $-1.4$ to 1.1.
 Clip levels of {\emph I} maps: 1.7 mJy (8.9 GHz), 2.3 mJy (12.9 GHz).
 Beam FWHM and Position Angle: 0.95$\times$0.84 mas, $82.5^{\circ}$ (8.9 GHz).
 Contours: 8.9 GHz {\emph I}, starting at 1.70 mJy and increasing by factors of 2.
 (c) 15.4--43 GHz. Spectral index scale range: $-1$ to 0.6.
 Clip levels of {\emph I} maps: 1.8 mJy (15.4 GHz), 7.7 mJy (43 GHz).
 Beam FWHM and Position Angle: 0.55$\times$0.50 mas, $83.6^{\circ}$ (15.4 GHz).
 Contours: 15.4 GHz {\emph I}, starting at 1.82 mJy and increasing by factors of 2.
 The arrows connect the plots of the spectral index slices to their corresponding maps. The hollow circle on the slice plot indicates the core position at the lower frequency of the pair.}
 \label{2007spix}
\end{figure*}

In Fig.~\ref{2007spix} we show the spectral index distribution of 2007+777 from 4.6--43 GHz. For this source we find a flat to inverted spectrum in the core region, falling off to a steep optically thin spectrum further downstream in the jet. The slices for the 4.6--7.9 GHz and 8.9--12.9 GHz maps (Fig.\ref{2007spix}a, b) show optically thick regions at the edge of the jet $\sim$8 mas from the core; however, these are possibly spurious features due to edge effects when combining the two maps. We only show the spectral index map from 15.4--43 GHz due to loss of flux in the 22-GHz map.

\begin{figure}
\includegraphics[width=90mm]{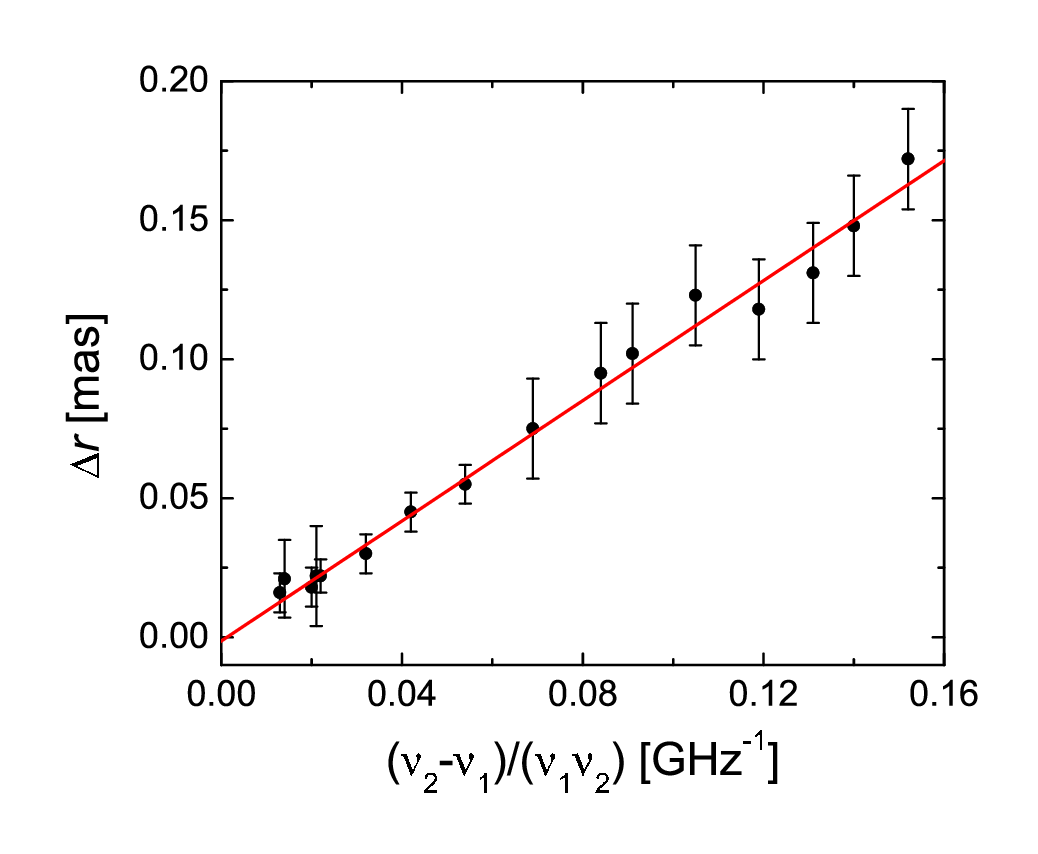}
 \caption{Straight-line fit to the all the core-shift data for 2007+777, assuming equipartition. The y-intercept value is equal to $-0.001\pm0.004$ mas. A straight-line fit through the origin is expected for $k_r=1$.}
 \label{all2007}
\end{figure}

\begin{figure}
\includegraphics[width=90mm]{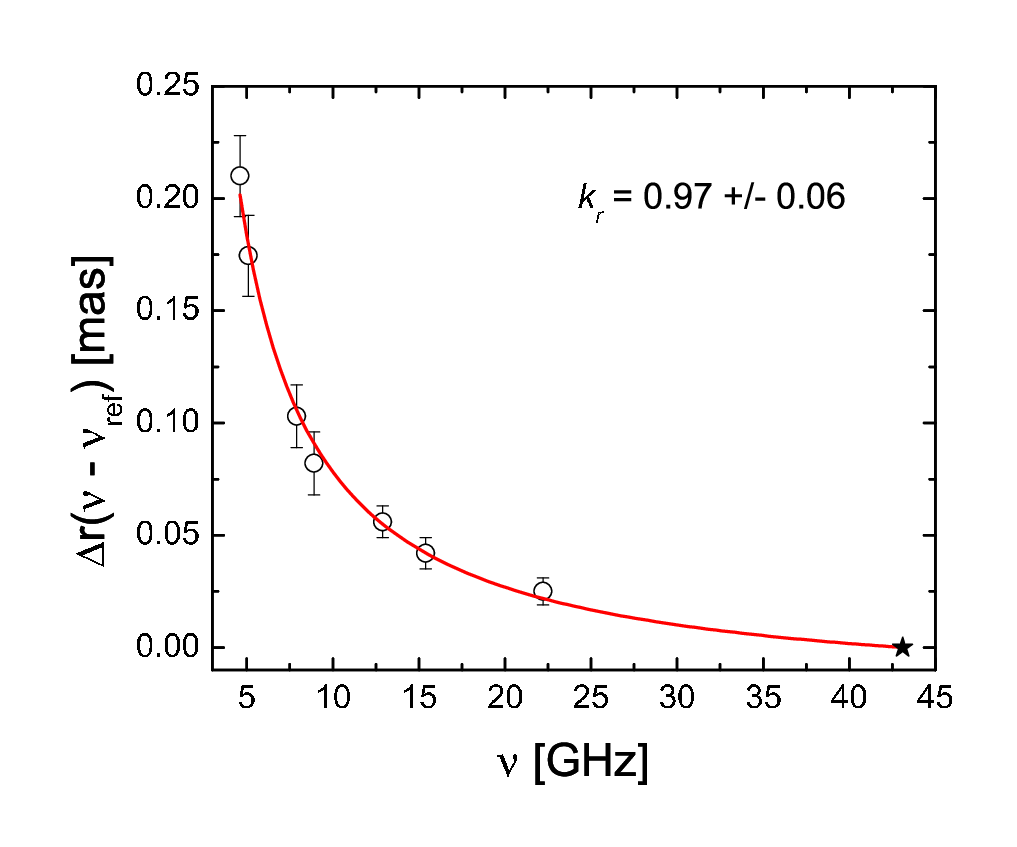}
 \caption{Plot of core-shift measurements versus frequency for 2007+777 (listed in Table \ref{43ref2007}), using 43 GHz as the reference frequency. Best-fit line: $\Delta r=A(\nu^{-1/k_r}-43.1^{-1/k_r})$ with parameters $A=1.08\pm0.11$ and $k_r=0.97\pm0.06$}
 \label{kr2007}
\end{figure}

All measured core-shifts for 2007+777 are listed in Table 5. The average direction of the measured shift is $89\pm13^{\circ}$, which is consistent with the parsec scale jet direction of $\sim-90^{\circ}$. The y-intercept of the best-fit line to the data (Fig.~\ref{all2007}) is $-0.001\pm0.004$, indicating that the compact jet region from 4.6--43 GHz is close to equipartition. Due to a failure to find shift measurements that produced reasonable spectral index maps combining the  higher frequencies with the 7.9 and 8.9 GHz images, we do not have as many measurements as in the previous case of 2200+420. In order to make the spectral index map between 8.9 and 12.9 GHz, we use in Fig.~\ref{2007spix} the characteristic position-offset measure of $5.2\pm0.3$ ${\rm pc}\cdot{\rm GHz}$ from the best-fit line to the measured data in Fig.~\ref{all2007} to estimate the required shift. This produces a spectal index map consistent with those generated at lower and higher frequencies (Fig.~\ref{2007spix}). In order to find an estimate for the parameter $k_r$, we use over-lapping shifts to plot 7.9--43 GHz and 8.9--43 GHz values (e.g., (4.6--15.4 GHz)$-$(4.6--7.9 GHz)$+$(15.4--43 GHz)). Hence, we find $k_r=0.97\pm0.06$ (Fig.~\ref{kr2007}) for this source (see Table 6 for listed values). Without the 7.9--43 GHz and 8.9--43 GHz points, we get $k_r=0.99\pm0.05$. 

\cite{gabuzda1994} find $\beta_{{\rm app}}=2.9$ for 2007+777, where $\beta_{\rm app}$ is the apparent velocity of the jet in units of the speed of light, based on three observations between 1980 and 1989. However, from much more densely sampled multi-epoch monitoring from 1994--2002 at 15~GHz, \cite{kellermann2004} do not detect any outward motion from the jet of 2007+777. \cite{hovatta2009} derive a Doppler factor ($\delta$) of 7.9 using variability arguments from single dish monitoring indicating that this jet does have a relativistic flow. To estimate the magnetic field strength in this source, we require estimates for the bulk Lorentz factor ($\Gamma$) and the viewing angle ($\theta$). Due to the uncertainty in the jet speed measurements, we assume $\beta_{\rm app}\sim\delta$ and from this we find the minimum Lorentz factor of the flow from $\Gamma_{\rm min}=(1+\beta_{\rm app}^2)^{1/2}$. To estimate the viewing angle we use $\theta=\sin^{-1}(1/\Gamma_{\rm min})$; while this in unlikely to be true for any one source, it is a good approximation in a statistical sense \citep[see][]{cohen}.
We find a jet opening angle of 1.0$^{\circ}$ by estimating the projected half opening angle using the ratio between the apparent transverse size and longitudinal distance of the jet components within 1 mas from the core (i.e., we assume a conical jet with a constant opening angle within this region, see also \cite{jorstad2005}).

Due to the absence of direct jet speed measurements, the following magnetic field values can only be considered as rough estimates. Using the same procedure as described in the case of 2200+420, the calculated $B_{core}$, $r_{core}$ and $B_1$ are presented in Table \ref{2007results} using the equipartition value of $k_r=1$ and a jet spectral index of $-0.5$ consistent with the spectral index maps (Fig.~\ref{2007spix}). 
Considering a range of values for $k_r$, from the fit to the data shown in Fig.~\ref{kr2007}, of $0.91<k_r<1.03$, we get $8.7<r_{core}(43{\rm~GHz})<0.5$ pc and $0.23<B_1<0.27$ pc (using $\Omega_{r\nu}$ at 43~GHz).

\begin{figure}
\includegraphics[width=100mm]{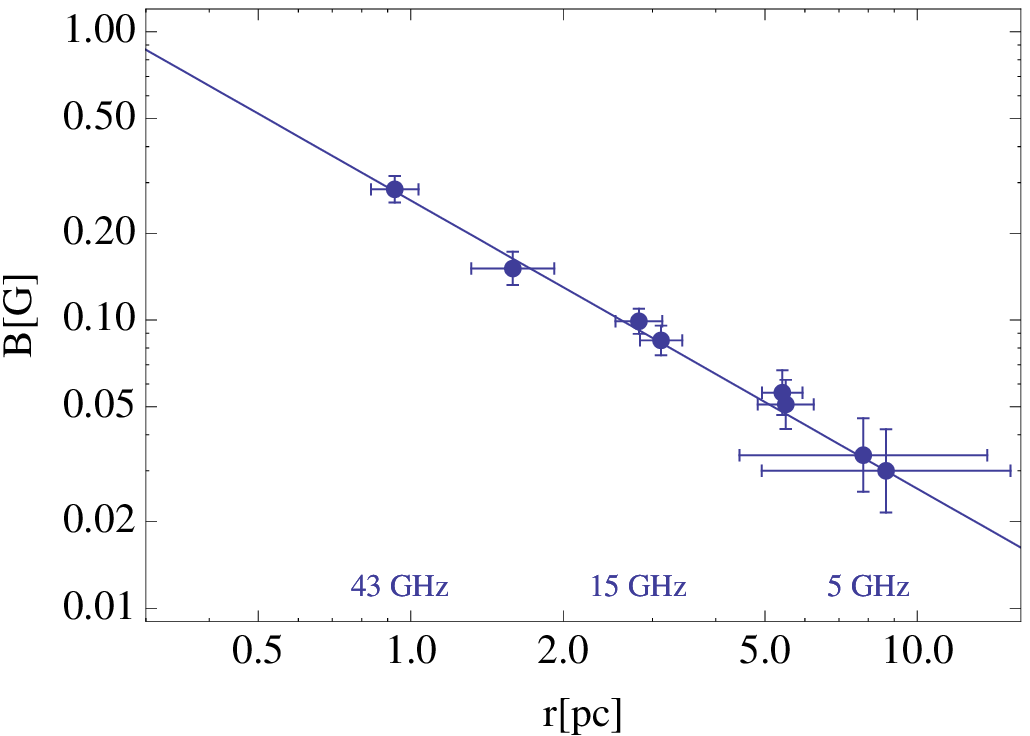}
 \caption{Plot of magnetic field strength ($B$) in units of Gauss versus distance along the jet ($r$) in units of parsecs for the core of 2007+777 at 4.6, 5.1, 7.9, 8.9, 12.9, 15.4, 22.2 \& 43 GHz using $k_r=1$. Also plotted is the best-fit line $B=B_1r^{-1}$ with $B_1=0.257$ G with a formal error of 0.007 G.}
 \label{bvr2007}
\end{figure}

\begin{figure}
\includegraphics[width=100mm]{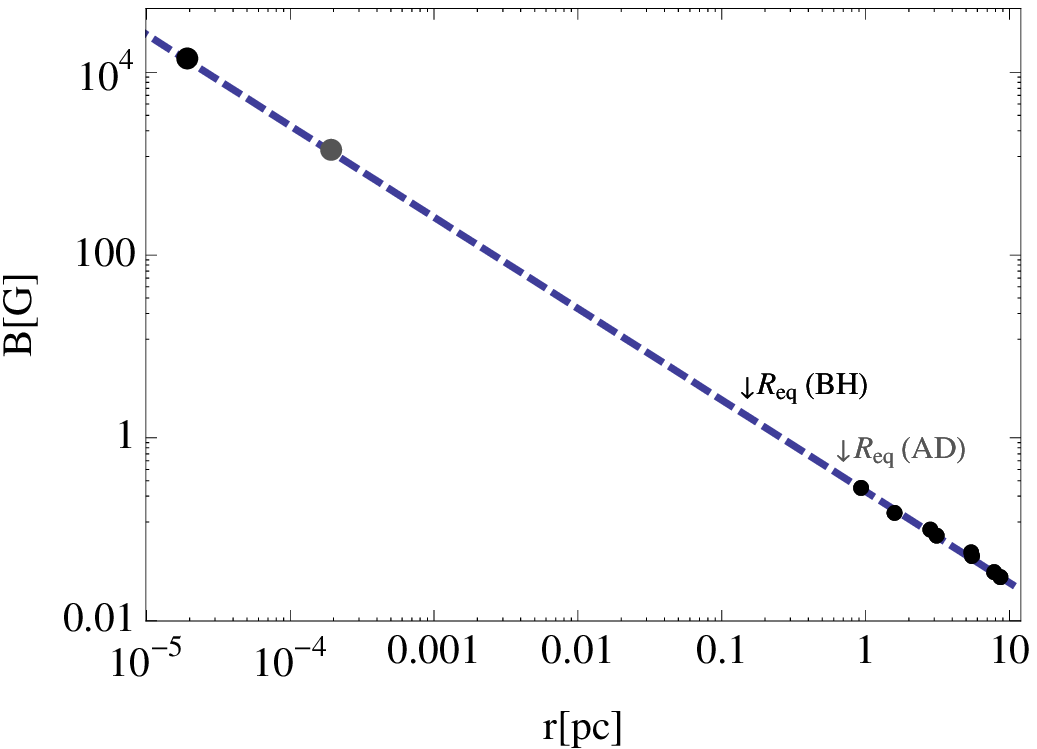}
 \caption{Extended plot of $B$ versus $r$ for 2007+777 to include the extrapolated magnetic field strength ($B\simeq0.26r^{-1}$) at the accretion disk (10$r_g$) and black holes ($r_g$) distances. The estimated equipartition radii from the model K07 are indicated by the arrows with $R_{\rm eq,AD}\sim0.7$ pc and $R_{\rm eq,BH}\sim0.1$ pc. Also plotted are points at 10$r_g$ (grey) and $r_g$ (black) representing the theoretically expected magnetic field strengths for a magnetically powered jet.}
 \label{bvrext2007}
\end{figure}

We plot $B_{core}$ versus $r_{core}$, for $k_r=1$, in Fig.~\ref{bvr2007} and find $B_1=0.257\pm0.007$ G. Extending this line to the launching region of the jet (Fig.~\ref{bvrext2007}), we find that it reaches $\sim1.3\times10^{3}$ G at an accretion disk distance of 10$r_g$ and $\sim1.3\times10^{4}$ G at $r_g$, using a black hole mass of $4.0\times10^8 M_\odot$ \citep{jiang}. We calculate the total jet power using $B_1\simeq0.26$ G and with the jet parameters listed in Table~\ref{source2200} to find $L_{\rm total}\sim(3.2\pm0.2)\times10^{45}$ erg s$^{-1}$. Using   $L_{\rm total}$, we estimate the magnetic field strength of $B_0\sim1.4\times10^{3}$ G at the jet launching distance of $10r_g$, which is in excellent agreement with our extrapolated magnetic field strength results (Fig.~\ref{bvrext2007}).

The estimated equipartition radius for an accretion-disk-launched jet of $\sim0.7$ pc is just outside the range of 43-GHz VLBI, which probes distances of $\sim0.9$ pc from the central engine (Fig.~\ref{bvrext2007}). However, this source should be a good candidate for the possible detection of accelerating jet components from multi-epoch monitoring at 43~GHz, even though the 43-GHz core may extend as far as $\sim9$ pc from the base of the jet, since the jet flow is not expected to reach its terminal Lorentz factor until $\sim1$ pc for a black-hole-launched jet and $\sim7$ pc for an accretion-disk-launched jet. 

\begin{table}
 \caption{All measured core-shifts for 2007+777, plotted in Fig.~\ref{all2007}}
 \label{shifts2007}
 \begin{tabular}{@{}cccccccc}
  \hline
  $\nu_1$ & $\nu_2$ & $\Delta r$ &   Angle         \\
    (GHz)  &  (GHz)    &   (mas)     &   ($^{\circ}$)  \\
  \hline
  4.6 & 5.1 & $0.02\pm0.02$ & $91\pm1$  \\
  4.6 & 7.9 & $0.10\pm0.02$ & $89\pm1$  \\ 
  4.6 & 8.9 & $0.12\pm0.02$ & $89\pm1$  \\
  4.6 & 12.9 & $0.15\pm0.02$ & $79\pm1$  \\
  4.6 & 15.4 & $0.17\pm0.02$ & $87\pm1$  \\
  5.1 & 7.9 & $0.08\pm0.02$ & $97\pm1$  \\
  5.1 & 8.9 & $0.10\pm0.02$ & $88\pm1$  \\
  5.1 & 12.9 & $0.12\pm0.02$ & $82\pm1$  \\
  5.1 & 15.4 & $0.13\pm0.02$ & $92\pm1$  \\
  7.9 & 8.9 & $0.02\pm0.01$ & $91.2\pm0.8$  \\
  12.9 & 15.4 & $0.016\pm0.007$ & $117.1\pm0.3$ \\
  12.9 & 22.2 & $0.030\pm0.007$ & $104.1\pm0.4$ \\
   12.9 & 43.1 & $0.055\pm0.007$ & $84.4\pm0.4$ \\
  15.4 & 22.2 & $0.018\pm0.007$ & $54.8\pm0.3$ \\
  15.4 & 43.1 & $0.045\pm0.007$ & $86.7\pm0.4$ \\
  22.2 & 43.1 & $0.022\pm0.006$ & $83.5\pm0.3$ \\
  \hline
 \end{tabular}
\\Col.~(1) \& (2): observed frequency in GHz, Col.~(3): magnitude of the core-shift in milliarcseconds, Col.~(4): angle of core-shift in degrees. Average angle: $89\pm13^{\circ}$
\end{table}

\begin{table}
 \caption{Averaged core-shifts using 43 GHz as the reference frequency for 2007+777, plotted in Fig.~\ref{kr2007}}
 \label{43ref2007}
 \begin{tabular}{@{}cccc}
  \hline
  $\nu$ & $\Delta r$ &  $\Delta r_{projected}$ & Fraction of beam \\
   (GHz) &   (mas)    &       (pc)       &   (\%)     \\
  \hline
  4.6 & $0.21\pm0.02$ & $1.0\pm0.1$ & 9\\
  5.1 & $0.17\pm0.02$  & $0.8\pm0.1$ & 8\\
  7.9 & $0.10\pm0.02$  & $0.5\pm0.1$ & 7\\
  8.9 & $0.08\pm0.02$  & $0.4\pm0.1$ & 6\\
  12.9 & $0.056\pm0.007$  & $0.27\pm0.03$ & 6\\
  15.4 & $0.042\pm0.007$  & $0.20\pm0.03$ & 6\\
  22.2 & $0.025\pm0.006$  & $0.12\pm0.03$ & 5\\
  \hline
 \end{tabular}
\\Col.~(1): observed frequency in GHz, Col.~(2): magnitude of the shift from the 43-GHz core position, Col.~(3): projected shift in parsecs (1~mas = 4.8 pc), Col.~(4): core-shift as a fraction of the beam-size at the corresponding frequency, in percent.
\end{table}

\begin{table}
 \caption{Derived physical quantities for 2007+777 ($k_r=1$)}
 \label{2007results}
 \begin{tabular}{@{}cccccccc}
  \hline
  $\nu$ & $B_{core}(\nu)$ & $r_{core}(\nu)$ & $B_{1\rm{pc}}$ & $\Omega_{r \nu}$\\
   (GHz) &       (G)             &        (pc)            &          (G)        &(${\rm pc} \cdot {\rm GHz}$)\\
  \hline
  4.6    & $0.030\pm0.010$     & $8.7\pm4.9$    & $0.26\pm0.22$     & -\\
  5.1    & $0.034\pm0.010$     & $7.8\pm4.4$    & $0.26\pm0.22$     & $5.0\pm2.8$\\
  7.9    & $0.051\pm0.003$   & $5.5\pm0.7$    & $0.28\pm0.05$    & $5.5\pm0.7$\\
  8.9    & $0.056\pm0.003$   & $5.4\pm0.5$    & $0.30\pm0.05$    & $6.0\pm0.6$\\
  12.9   & $0.085\pm0.005$  & $3.1\pm0.3$    & $0.26\pm0.04$    & $5.0\pm0.5$\\
  15.4   & $0.099\pm0.005$  & $2.8\pm0.3$    & $0.28\pm0.04$   & $5.5\pm0.5$\\
  22.2   & $0.15\pm0.01$     & $1.6\pm0.3$    & $0.24\pm0.06$    & $4.4\pm0.7$\\
  43.1   & $0.28\pm0.03$     & $0.9\pm0.2$    & $0.26\pm0.07$    & $5.0\pm0.9$\\
  \hline
 \end{tabular}
\\Col.~(1): observed frequency in GHz, Col.~(2): radio core magnetic field strength in Gauss, Col.~(3): de-projected distance of the radio core to the base of the jet in parsecs, Col.~(4): magnetic field strength at 1 pc, Col.~(5): average core-position offset measure, in units of $10^9 {\rm pc} \cdot {\rm GHz}$, for shifts with $\nu_2$ in Table~\ref{shifts2007} equal to $\nu$ in Col.~(1) above.
Note: Physical parameters at 4.6~GHz calculated with 5.1 GHz core-position offset measure.
 \end{table}


\subsection{1418+546}
\begin{figure*}
\includegraphics[width=130mm]{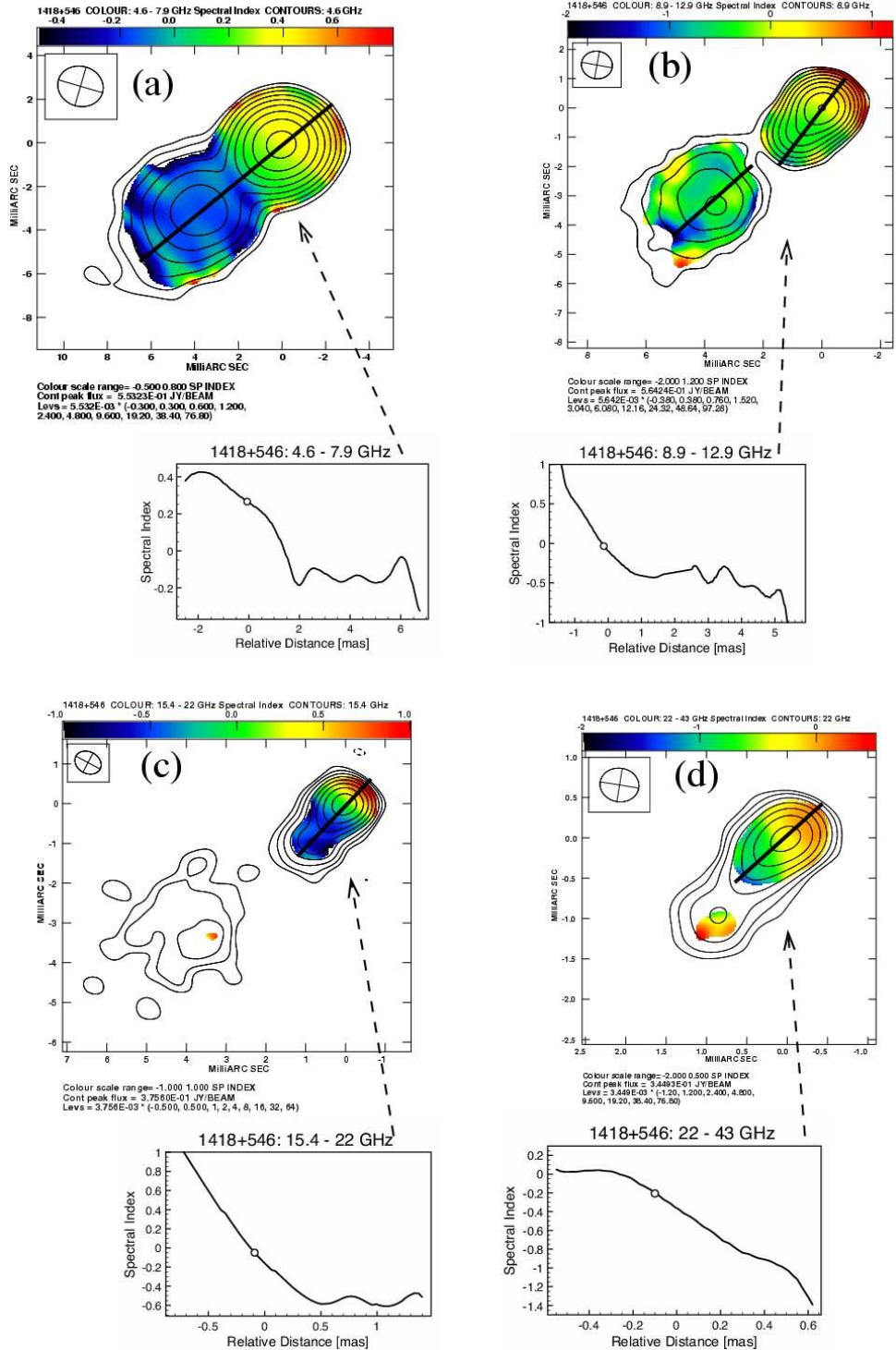}
 \caption{Spectral index maps for 1418+546 constructed from frequency pairs, with plots of slices along the jet region indicated by the solid line.
 (a) 4.6--7.9 GHz. Spectral index scale range: $-0.5$ to 0.8.
 Clip levels of {\emph I} maps: 1.7 mJy (4.6 GHz), 2.6 mJy (7.9 GHz).
 Beam FWHM and Position Angle: 2.10$\times$1.84 mas, $73.6^{\circ}$ (4.6 GHz).
 Contours: 4.6 GHz {\emph I}, starting at 1.66 mJy and increasing by factors of 2.
 (b) 8.9--12.9 GHz. Spectral index scale range: $-2$ to 1.2.
 Clip levels of {\emph I} maps: 2.1 mJy (8.9 GHz), 3.7 mJy (12.9 GHz).
 Beam FWHM and Position Angle: 1.06$\times$0.94 mas, $79.6^{\circ}$ (8.9 GHz).
 Contours: 8.9 GHz {\emph I}, starting at 2.14 mJy and increasing by factors of 2.
 (c) 15.4--22 GHz. Spectral index scale range: $-1.0$ to 1.0.
 Clip levels of {\emph I} maps: 3.0 mJy (15.4 GHz), 5.8 mJy (22 GHz).
 Beam FWHM and Position Angle: 0.69$\times$0.54 mas, $61.9^{\circ}$ (15.4 GHz).
 Contours: 15.4 GHz {\emph I}, starting at 1.88 mJy and increasing by factors of 2.
 (d) 22--43 GHz. Spectral index scale range: $-2.0$ to 0.5.
 Clip levels of {\emph I} maps: 4.2 mJy (22 GHz), 9.2 mJy (43 GHz).
 Beam FWHM and Position Angle: 0.48$\times$0.42 mas, $80.8^{\circ}$ (22 GHz).
 Contours: 22 GHz {\emph I}, starting at 4.14 mJy and increasing by factors of 2.
 The arrows connect the plots of the spectral index slices to their corresponding maps. The hollow circle on the slice plot indicates the core position at the lower frequency of the pair.}
 \label{1418spix}
\end{figure*}

Fig.~\ref{1418spix} shows the spectral index distribution from 4.6--43 GHz for the source 1418+546. We find an inverted core spectrum between the lower frequency pairs (Fig.~\ref{1418spix}a, b, c) with a flat spectrum core region between 22 and 43 GHz (Fig.~\ref{1418spix}d). In the 4.6--7.9 GHz map (Fig.~\ref{1418spix}a) the jet spectral index is $\sim$-0.2, while at higher frequencies it becomes steeper with a typical value of $-0.5$ (Fig.~\ref{1418spix}b, c). 

The best fit line to the measured shifts listed in Table 8 gives a y-intercept of $0.003\pm0.004$ assuming $k_r=1$ and the slope provides a characteristic $\Omega_{r\nu}=3.9\pm0.2$ ${\rm pc}\cdot{\rm GHz}$ (Fig.~12). The average position angle of the shift is $-52\pm13^{\circ}$, consistent within the errors with the inner jet direction at 43 GHz of $\sim137^{\circ}$. Again averaging the measured shifts and using 43 GHz as the reference frequency (Table 9), we find $k_r=1.08\pm0.11$ between 4.6 and 43 GHz (Fig.~\ref{kr1418}). The equipartition value $k_r=1$ and $\alpha_j=-0.5$ are used to calculate the physical parameters listed in Table 10. We estimate a jet opening angle of $1.9^{\circ}$ from the size of the jet components within 1 mas from the core. We use a Doppler factor of 5.1 from \cite{hovatta2009} and $\beta_{{\rm app}}=4.3$ from \cite{pushkarev1999} to find estimates of the bulk Lorentz factor of 4.5 using $\Gamma=(\beta_{{\rm app}}^{2}+\delta^2+1)/(2\delta)$ and the jet viewing angle of 11.2$^{\circ}$ using $\theta=\tan^{-1}((2\beta_{{\rm app}})/((\beta_{{\rm app}}^{2}+\delta^2-1))$ \citep[e.g.,][]{hovatta2009}. 

We find $B_1=0.178\pm0.006$ G from our plot of $B_{core}$ versus $r_{core}$ (Fig.~\ref{bvr1418}). Extrapolating this line, we find $B\sim680$ G at an accretion disk distance of 10$r_g$, using the black hole mass estimate of 10$^{8.74}$ from \cite{jiang}. Using the calculated total luminosity of $(1.7\pm0.1)\times10^{45}$ erg s$^{-1}$, we expect $B_0\sim750$ G at $10r_g$ from the divergence theorem (Fig.~\ref{bvrext1418}). 

For $0.97<k_r<1.19$, we get $1.1<r_{core}(43{\rm~GHz})<0.01$ pc and $0.18<B_1<0.21$ pc (using $\Omega_{r\nu}$ at 43~GHz). An interesting feature of this source is that the estimated equipartition radius for an accretion-disk-launched jet ($R_{eq}\sim0.5$ pc) is within the range of VLBI observations at 43 GHz assuming $k_r\sim1$. 
Multi-epoch VLBI monitoring of this jet at 43 GHz should provide evidence for the extended acceleration predicted from the magnetically driven jet model of K07 since the jet flow is not expected to reach its terminal Lorentz factor until a distance of $\sim1$ pc for a black-hole-launched jet and $\sim5$ pc for an accretion-disk-launched jet. 

\begin{table}
 \caption{All measured core-shifts for 1418+546, plotted in Fig.~\ref{all1418}}
 \label{shifts1418}
 \begin{tabular}{@{}cccccccc}
  \hline
  $\nu_1$ & $\nu_2$ & $\Delta r$ &   Angle          \\
    (GHz)  &  (GHz)    &   (mas)     &   ($^{\circ}$) \\
  \hline
  4.6 & 5.1 & $0.03\pm0.02$ & $-38.2\pm0.9$  \\
  4.6 & 7.9 & $0.14\pm0.02$ & $-49.6\pm0.7$  \\ 
  4.6 & 8.9 & $0.15\pm0.02$ & $-44.0\pm0.6$  \\
  4.6 & 12.9 & $0.20\pm0.02$ & $-51.7\pm0.7$  \\
  4.6 & 15.4 & $0.23\pm0.02$ & $-60.3\pm0.5$  \\
  5.1 & 7.9 & $0.11\pm0.02$ & $-41.5\pm0.8$  \\
  5.1 & 8.9 & $0.13\pm0.02$ & $-35.8\pm0.6$  \\
  5.1 & 12.9 & $0.17\pm0.02$ & $-46.8\pm0.4$  \\
  5.1 & 15.4 & $0.21\pm0.02$ & $-54.0\pm0.5$  \\
  7.9 & 8.9 & $0.02\pm0.02$ & $-24.3\pm0.7$  \\
  7.9 & 12.9 & $0.07\pm0.02$ & $-53.2\pm0.7$  \\
  7.9 & 15.4 & $0.11\pm0.02$ & $-57.4\pm0.2$  \\
  8.9 & 12.9 & $0.06\pm0.02$ & $-58.4\pm0.9$  \\
  8.9 & 15.4 & $0.09\pm0.02$ & $-68.3\pm0.7$  \\
  12.9 & 15.4 & $0.021\pm0.007$ & $-82.4\pm0.3$  \\
  15.4 & 22.2 & $0.030\pm0.007$ & $-62.7\pm0.3$  \\
  22.2 & 43.1 & $0.036\pm0.005$ & $-56.2\pm0.2$  \\
  \hline
 \end{tabular}
\\Col.~(1) \& (2): observed frequency in GHz, Col.~(3): magnitude of the core-shift in milliarcseconds, Col.~(4): angle of core-shift in degrees. Average angle: $-52\pm13^{\circ}$
\end{table}

\begin{table}
 \caption{Averaged core-shifts using 43 GHz as reference frequency for 1418+546, plotted in Fig.~\ref{kr1418}}
 \label{kr1418}
 \begin{tabular}{@{}cccc}
  \hline
  $\nu$ & $\Delta r$ &  $\Delta r_{projected}$ & Fraction of beam \\
   (GHz) &   (mas)    &       (pc)       &   (\%)     \\
  \hline
  4.6 & $0.29\pm0.02$ & $0.75\pm0.05$ & 10\\
  5.1 & $0.27\pm0.02$  & $0.70\pm0.05$ & 10\\
  7.9 & $0.16\pm0.02$  & $0.42\pm0.05$ & 9\\
  8.9 & $0.13\pm0.02$  & $0.34\pm0.05$ & 9\\
  12.9 & $0.085\pm0.007$  & $0.22\pm0.02$ & 8\\
  15.4 & $0.066\pm0.007$  & $0.17\pm0.02$ & 8\\
  22.2 & $0.036\pm0.005$  & $0.09\pm0.01$ & 6\\
  \hline
 \end{tabular}
\\Col.~(1): observed frequency in GHz, Col.~(2): magnitude of the shift from the 43-GHz core position, Col.~(3): projected shift in parsecs (1~mas = 2.6 pc), Col.~(4): core-shift as a fraction of the beam-size at the corresponding frequency, in percent.
\end{table}

\begin{table}
 \caption{Derived physical quantities for 1418+546 ($k_r=1$)}
 \label{1418results}
 \begin{tabular}{@{}cccccccc}
  \hline
  $\nu$ & $B_{core}(\nu)$ & $r_{core}(\nu)$ & $B_{1\rm{pc}}$ & $\Omega_{r \nu}$\\
   (GHz) &       (G)             &        (pc)            &          (G)        &(${\rm pc} \cdot {\rm GHz}$)\\
  \hline
  4.6    & $0.040\pm0.011$     & $4.1\pm2.3$    & $0.16\pm0.14$     & -\\
  5.1    & $0.044\pm0.012$     & $3.7\pm2.1$    & $0.16\pm0.14$     & $3.7\pm2.0$\\
  7.9    & $0.067\pm0.004$   & $2.7\pm0.3$    & $0.18\pm0.03$    & $4.1\pm0.4$\\
  8.9    & $0.076\pm0.004$   & $2.2\pm0.2$    & $0.17\pm0.02$    & $3.8\pm0.4$\\
  12.9   & $0.11\pm0.01$  & $1.6\pm0.3$    & $0.17\pm0.06$    & $3.9\pm0.8$\\
  15.4   & $0.13\pm0.01$  & $1.5\pm0.2$    & $0.19\pm0.03$   & $4.4\pm0.5$\\
  22.2   & $0.19\pm0.02$     & $0.9\pm0.2$    & $0.17\pm0.04$    & $4.0\pm0.7$\\
  43.1   & $0.36\pm0.02$     & $0.52\pm0.05$    & $0.19\pm0.03$    & $4.3\pm0.4$\\
  \hline
 \end{tabular}
\\Col.~(1): observed frequency in GHz, Col.~(2): radio core magnetic field strength in Gauss, Col.~(3): de-projected distance of the radio core to the base of the jet in parsecs, Col.~(4): magnetic field strength at 1 pc, Col.~(5): average core-position offset measure, in units of $10^9 {\rm pc} \cdot {\rm GHz}$, for shifts with $\nu_2$ in Table~\ref{shifts1418} equal to $\nu$ in Col.~(1) above.
Note: Physical parameters at 4.6~GHz calculated with 5.1 GHz core-position offset measure.
 \end{table}

\subsection{0954+658}
In Fig.~\ref{0954spix} we show the spectral index maps for 0954+658 from 4.6--43 GHz. We observe a strongly inverted spectrum from 4.6--22 GHz in the core region (Fig.~\ref{0954spix}a, b, c), which then flattens considerably at the highest frequencies (Fig.~\ref{0954spix}d). We note the spectral index gradient across the jet in the 4.6--7.9 GHz image, which is discussed in \cite{osullivangabuzda2009}. The region of inverted spectrum at $\sim$3 mas from the core in the 8.9--12.9 GHz map (Fig.~\ref{0954spix}b) may also be connected with the change in the jet direction in the plane of the sky. The extended emission in this source falls off sharply with increasing frequency, which is why we could not obtain core-shift measurements at frequencies higher than 8.9 GHz.

We obtained two reliable core-shift measurements for the jet of 0954+658, with the angle of the shifts ($\sim127^{\circ}$) consistent with the inner jet direction at 43 GHz of $\sim-59^{\circ}$ (Table 11). 
In this case, we assume a value of $k_r=1$ from 4.6--43 GHz due to our lack of measurements, and calculate the shifts with which to align our images from this. The spectral index maps in Fig.~\ref{0954spix} indicate that this is a reasonable assumption between 4.6 and 43 GHz, providing a flat to inverted core spectrum with a steep spectrum jet as expected.

In order to estimate the magnetic field strength for this source, we use a Doppler factor of 6.2 from \cite{hovatta2009} and $\beta_{{\rm app}}=6.2$ from \cite{gabuzda1994} to find $\Gamma=6.3$ and $\theta=9.3^{\circ}$. We get $\phi=2.2^{\circ}$ from the size of jet components within 1 mas of the core. We use a jet spectal index of -0.6 throughout the calculations due to the steep fall-off in the jet spectral index.
In this case, to calculate the magnetic field strength we need to assume a value for the lower cut-off in the electron energy distribution since $B_1\propto\gamma_{\rm min}^{-0.05}$. Using $\Omega_{r\nu}=8.8\pm0.7$ ${\rm pc}\cdot{\rm GHz}$, we get $B_1=0.27\pm0.03$ G with $\gamma_{\rm min}=100$ while if $\gamma_{\rm min}$ extends all the way down to 1, we get $B_1=0.34\pm0.05$ G.
Our estimates of the physical parameters for this source are shown in Table \ref{source0954} assuming $\gamma_{\rm min}=100$ and $\alpha=-0.6$.

Extrapolating these measurements, we estimate a magnetic field strength in the 43 GHz radio core of $B_{core}\sim0.21$ G at a distance of $r_{core}\sim1.3$ pc. Using a black hole mass of $10^{8.37}$ M$_\odot$ \citep{jiang}, we find $R_{\rm eq,AD}\sim0.2$ pc and $R_{eq}(BH)\sim0.02$ pc. 
From the calculated total luminosity of $(1.1\pm0.2)\times10^{46}$ erg s$^{-1}$, we expect $B_0\sim4\times10^3$ from the divergence theorem. Using $B(r)=0.27r^{-1}$, we find $B\sim2\times10^{3}$ G at 10$r_g$. 

\begin{table}
 \caption{Measured core-shifts for 0954+658}
 \label{shifts0954}
 \begin{tabular}{@{}cccccccc}
  \hline
  $\nu_1$ & $\nu_2$ & $\Delta r$ &   Angle         & fraction of beam \\
    (GHz)  &  (GHz)    &   (mas)     &   ($^{\circ}$) &          (\%)            \\
  \hline
  4.6 & 7.9 & $0.16\pm0.02$ & $126.8\pm0.6$ & 6 \\ 
  4.6 & 8.9 & $0.18\pm0.02$ & $127.3\pm0.6$ & 7 \\
    \hline
 \end{tabular}
\\Col.~(1) \& (2): observed frequency in GHz, Col.~(3): magnitude of the core-shift in milliarcseconds, Col.~(4): angle of core-shift in degrees. Col.~(5): core-shift as a fraction of the 4.6-GHz beam-size, in percent.
\end{table}

\begin{table}
 \caption{Derived physical quantities for 0954+658}
 \label{source0954}
 \begin{tabular}{@{}cccccccc}
  \hline
  $\nu$ & $B_{core}(\nu)$ & $r_{core}(\nu)$ & $B_{1\rm{pc}}$  & $\Omega_{r \nu}$\\
   (GHz) &       (G)             &        (pc)            &          (G)            &(${\rm pc} \cdot {\rm GHz}$)\\
  \hline
  7.9    & $0.038\pm0.002$     & $7.0\pm0.6$    & $0.27\pm0.03$      & $8.9\pm0.7$\\
  8.9    & $0.044\pm0.002$     & $6.0\pm0.4$    & $0.27\pm0.03$      & $8.7\pm0.6$\\
  \hline
 \end{tabular}
\\Col.~(1): observed frequency in GHz, Col.~(2): radio core magnetic field strength in Gauss, Col.~(3): de-projected distance of the radio core to the base of the jet in parsecs, Col.~(4): magnetic field strength at 1 pc, Col.~(5): average core-position offset measure, in units of $10^9 {\rm pc} \cdot {\rm GHz}$, for shifts with $\nu_2$ in Table~\ref{shifts0954} equal to $\nu$ in Col.~(1) above. Note: magnetic field strengths calculated for $\gamma_{\rm min}=100$
 \end{table}

\subsection{1156+295 and 1749+096}
For these two sources, we were unable to obtain core-shift measurements due to the weak nature of the jet emission. Hence, in this case we cannot be sure if the core corresponds to a stationary feature or the $\tau=1$ surface. To produce the spectral index maps (Figs.~\ref{1156spix},~\ref{1749spix}), we aligned the images by their model-fitted core positions. This produced uniform flat spectra in the core regions (e.g., the slice along the 4.6--7.9 GHz spectral index map for 1156+295, Fig.~\ref{1156spix}a). We can see that the core spectrum of 1156+295, is strongly inverted ($\alpha_{\rm 7.9 GHz}^{\rm 4.6 GHz}\sim1.3$) at low frequencies and then flattens considerably at the higher frequencies with $\alpha_{\rm 43 GHz}^{\rm 22 GHz}\sim-0.05$ (Fig.~\ref{1156spix}). The core spectrum of 1749+096 (Fig.~\ref{1749spix}) displays similar properties, with $\alpha_{\rm 7.9 GHz}^{\rm 4.6 GHz}\sim1.0$ and becomes flat at the highest frequencies with $\alpha_{\rm 43 GHz}^{\rm 22 GHz}\sim0.01$.

\section{Discussion}
In principle, the position of the vertex of the cone should provide a good estimate for the location of the central supermassive black hole powering the jet given that the distance to the radio core is likely to be much greater than the distance to the jet injection point or jet ``nozzle''. However, the exact position is uncertain due to the strong dependence of $r_{core}$ on small deviations in the value of $k_r$. In the case of 2200+420, the base of the jet can have a projected separation of up to $r_{\rm proj,max}\sim0.4$ mas ($\sim0.5$ pc for $k_r=0.92$) from the 43-GHz core. For 2007+777 and 1418+546, we find $r_{\rm proj,max}\sim0.2$ mas ($k_r=0.91$) and $\sim0.1$ mas ($k_r=0.97$), respectively, from their 43-GHz cores. Only the errors in the core-shift measurements are taken into account for our magnetic field strength calculations, hence, the results may be subject to systematic errors if one or more of the jet parameters ($\phi$, $\theta$, $\Gamma$, $\delta$) are inaccurate. Of course all values presented here should be considered in the context of the assumption that the jet speed measurements accurately reflect the bulk flow speed of the jet and not some pattern speed.


The main result from our core-shift measurements is that we find equipartition between the particle and magnetic field energy densities to be an accurate assumption for the radio core of the observed sources from 4.6--43 GHz. Conservation of particle number in a freely expanding conical jet requires $N\propto r^{-2}$, which then implies $B\propto r^{-1}$ from equipartition. Furthermore, we expect the longitudinal magnetic field strength to fall-off as $r^{-2}$ and the transverse component to fall-off as $r^{-1}$ \citep*{begelmanblandfordrees1984, huttermufson1986} so that over most of the length of the jet the transverse component will dominate and the overall magnetic field strength will scale as $r^{-1}$. This is also seen in models of magnetically dominated relativistic jets \citep{vlahakiskonigl2004} where, after the poloidal component of the magnetic field initially dominates, the toroidal component dominates at large distances and follows an approximate $r^{-1}$ fall-off in field strength. Our results are well described by $B\propto r^{-1}$, and by extrapolating back to the jet launching region, we get magnetic field strengths consistent with those required by models of magnetically launched jets to produce the observed jet luminosities. 

The implication of finding $N\propto r^{-2}$ consistent with our results is that, in the radio core from 4.6--43 GHz, continuous re-acceleration of the emitting particles is required to offset both the radiation and adiabatic expansion losses in this region. This could be achieved through numerous shocks \citep{marschergear1985}, magnetic reconnection \citep{zenitanihoshino2008} and/or turbulence from Kelvin-Helmholtz or current-driven instabilities.

In order to decouple the magnetic field strength from the particle number density, we used the equipartition condition of equation~\ref{equipartition}. We can now put the calculated $B_1$ back into the equation to get an estimate for the particle number density at 1 pc.
However, we also need to assume a value for the lower cut-off of the electron energy distribution ($\gamma_{\rm min}$). For a jet dominated by a normal (electron-proton) plasma, the energy distribution must have a cut-off of $\gamma_{\rm min}\leq100$ \citep{ghisellini1992, wardle1998}, while for an electron-positiron jet $\gamma_{\rm min}$ can extend all the way down to 1 \citep{reynolds1996, hirotani2005}. In Table~\ref{particledensity}, we list the particle number density in the jet at 1 parsec ($N_1$) for $\gamma_{\rm min}$ equal to 1 and to 100. 

Extended jet acceleration regions, most likely caused by MHD and/or hydrodynamical forces, have been detected in several AGN jets with the high resolution of VLBI \citep[e.g.,][]{jorstad2005}. From the model of K07, we can calculate the distance at which the jet approaches its terminal Lorentz factor. From this, we find that 2200+420, 2007+777 and 1418+546 would be interesting candidates for the detection of extended jet acceleration from VLBI monitoring, since the terminal Lorentz factor distance for accretion-disk-launched jets for these sources extends beyond the distance of the 43-GHz VLBI core from the base of the jet. \cite{jorstad2005} have already detected accelerating components from 43 GHz VLBA monitoring of 2200+420 providing more weight to the estimates presented in this paper. 

\begin{table}
 \caption{Particle number density estimates at 1 pc for each source ($k_r=1$).}
 \label{particledensity}
 \begin{tabular}{@{}cccc}
  \hline
  Source & $N_1(\gamma_{\rm min}=1)$ &  $N_1(\gamma_{\rm min}=100)$ \\
              &   (cm$^{-3}$)                               &      (cm$^{-3}$)      \\
  \hline
  0954+658 &          $1061$                         &     $6.8$               \\
  1156+295 &          $-$                                &     $-$               \\
  1418+546 &          $223$                           &     $2.2$               \\
  1749+096 &          $-$                                &     $-$               \\
  2007+777 &          $354$                           &     $3.5$               \\
  2200+420 &          $103$                           &     $1.0$               \\
  \hline
 \end{tabular}
\\Col.~(1): IAU source name, Col.~(2): particle number density at 1 pc in cm$^{-3}$ for a pair plasma with $\gamma_{\rm min}=1$, Col.~(3): particle number density at 1 pc in cm$^{-3}$ for a normal plasma with $\gamma_{\rm min}=100$.
\end{table}

Finally, we stress the importance of continued VLBI jet monitoring to obtain estimates of jet speeds, viewing angles and Doppler factors to enable magnetic field strength calculations for a much larger sample of AGN.

\section{Conclusions}

From our measurements of the frequency dependent position of the VLBI radio core, we find that:

1. The jets of 2200+420, 2007+777 and 1418+546 from 4.6--43 GHz are well described by the conical jet model of BK79 with $r_{\rm core}(\nu)\propto \nu^{-1/k_r}$. Furthermore, power-law fits to the data strongly support the equipartition value of $k_r=1$ in this region.

2. Our spectral index maps are also consistent with the BK79 jet model of an optically thick, unresolved jet base with all maps displaying a flat or inverted spectrum in the core region, turning over to an optically-thin steep-spectrum region further downstream in the jet.

3. We calculate equipartition magnetic field strengths with typical values from 10's to 100's of mG in the 4.6--43 GHz radio cores. Extrapolation of our results, following an $r^{-1}$ power law, to the accretion disk and black hole jet-launching distances gives magnetic field strengths in general agreement with those expected from theoretical models of magnetically powered jets.

\section*{Acknowledgements}
Funding for this research was provided by the Irish Research Council for Science, Engineering and Technology. The authors would like to thank Nektarios Vlahakis, Serguei Komissarov, Andrei Lobanov, Heino Falcke and Yuri Kovalev for their help as well as the anonymous referee for valuable comments that substantially improved this paper. The VLBA is a facility of the NRAO, operated by Associated Universities Inc. under cooperative agreement with the NSF. This research has made use of NASA's Astrophysics Data System Service. 

\bibliographystyle{mn2e}
\bibliography{osullivangabuzda_paper2}
\bsp

\begin{figure}
\includegraphics[width=90mm]{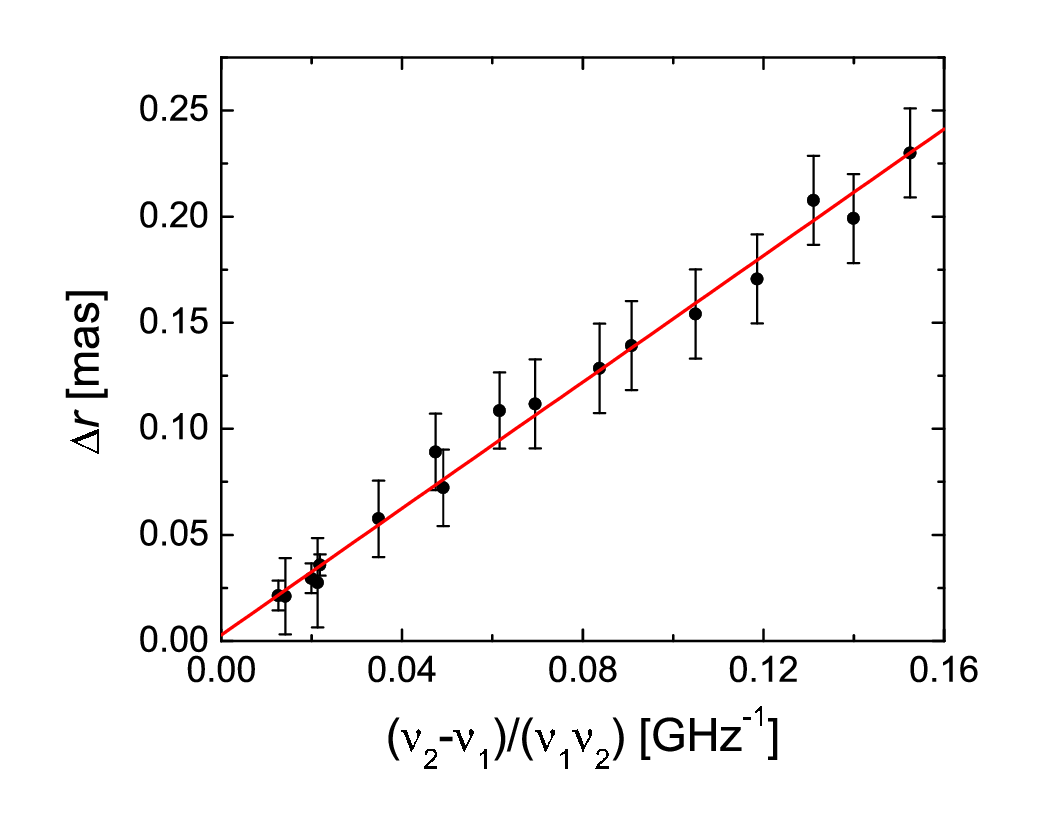}
 \caption{Straight-line fit to the all the core-shift data for 1418+546, assuming equipartition. The y-intercept value is equal to $0.003\pm0.004$ mas. A straight-line fit through the origin is expected for $k_r=1$.}
 \label{all1418}
\end{figure}

\begin{figure}
\includegraphics[width=90mm]{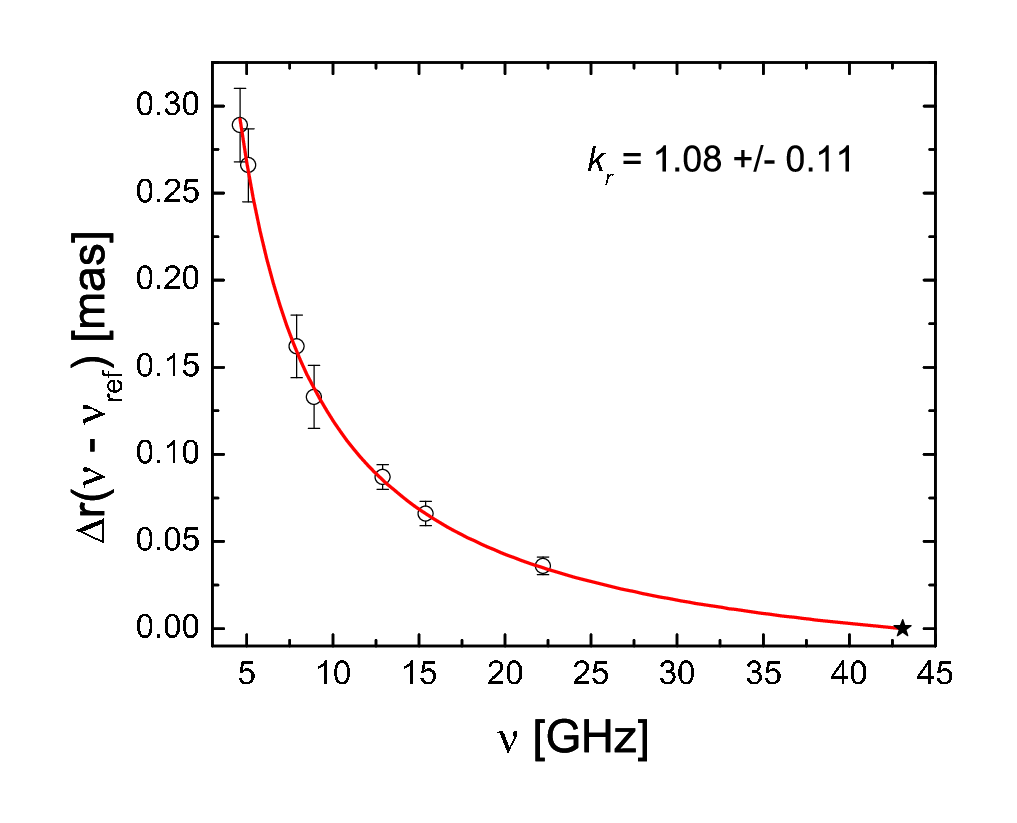}
 \caption{Plot of core-shift measurements versus frequency for 1418+546(listed in Table \ref{kr1418}), using 43 GHz as the reference frequency. Best-fit line: $\Delta r=A(\nu^{-1/k_r}-43.1^{-1/k_r})$ with parameters $A=1.4\pm0.2$ and $k_r=1.08\pm0.11$}
 \label{kr1418}
\end{figure}

\begin{figure}
\includegraphics[width=100mm]{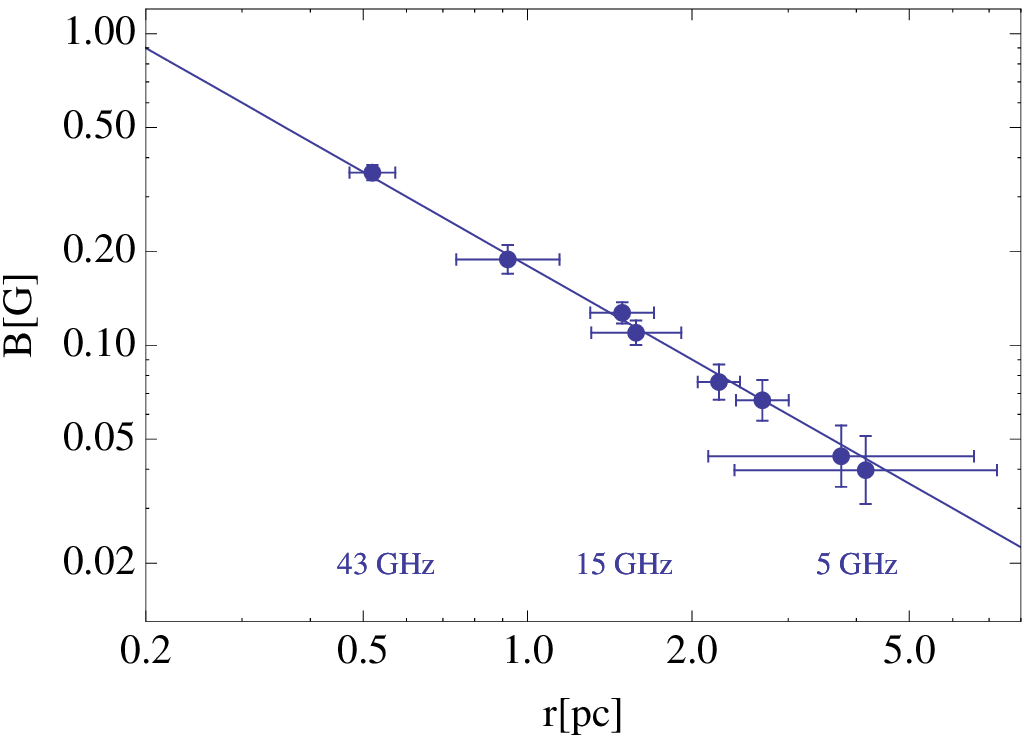}
 \caption{Plot of magnetic field strength ($B$) in units of Gauss versus distance along the jet ($r$) in units of parsecs for the core of 1418+546 at 4.6, 5.1, 7.9, 8.9, 12.9, 15.4, 22.2 \& 43 GHz using $k_r=1$. Also plotted is the best-fit line $B=B_1r^{-1}$ with $B_1=0.178$ G with a formal error of 0.006 G.}
 \label{bvr1418}
\end{figure}

\begin{figure}
\includegraphics[width=100mm]{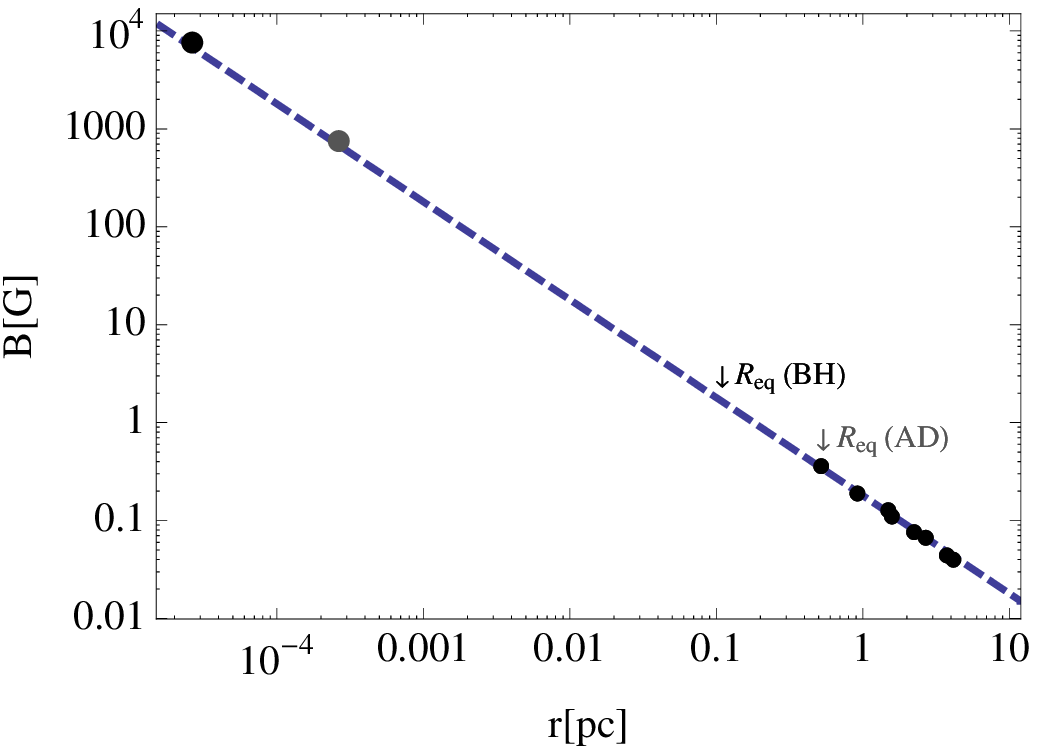}
 \caption{Extended plot of $B$ versus $r$ for 1418+546 to include the extrapolated magnetic field strength ($B\simeq0.18r^{-1}$) at the accretion disk (10$r_g$) and black holes ($r_g$) distances. The estimated equipartition radii from the model K07 are indicated by the arrows with $R_{\rm eq,AD}\sim0.5$ pc and $R_{\rm eq,BH}\sim0.1$ pc. Also plotted are points at 10$r_g$ (grey) and $r_g$ (black) representing the theoretically expected magnetic field strengths for a magnetically powered jet.}
 \label{bvrext1418}
\end{figure}

\begin{figure*}
\includegraphics[width=140mm]{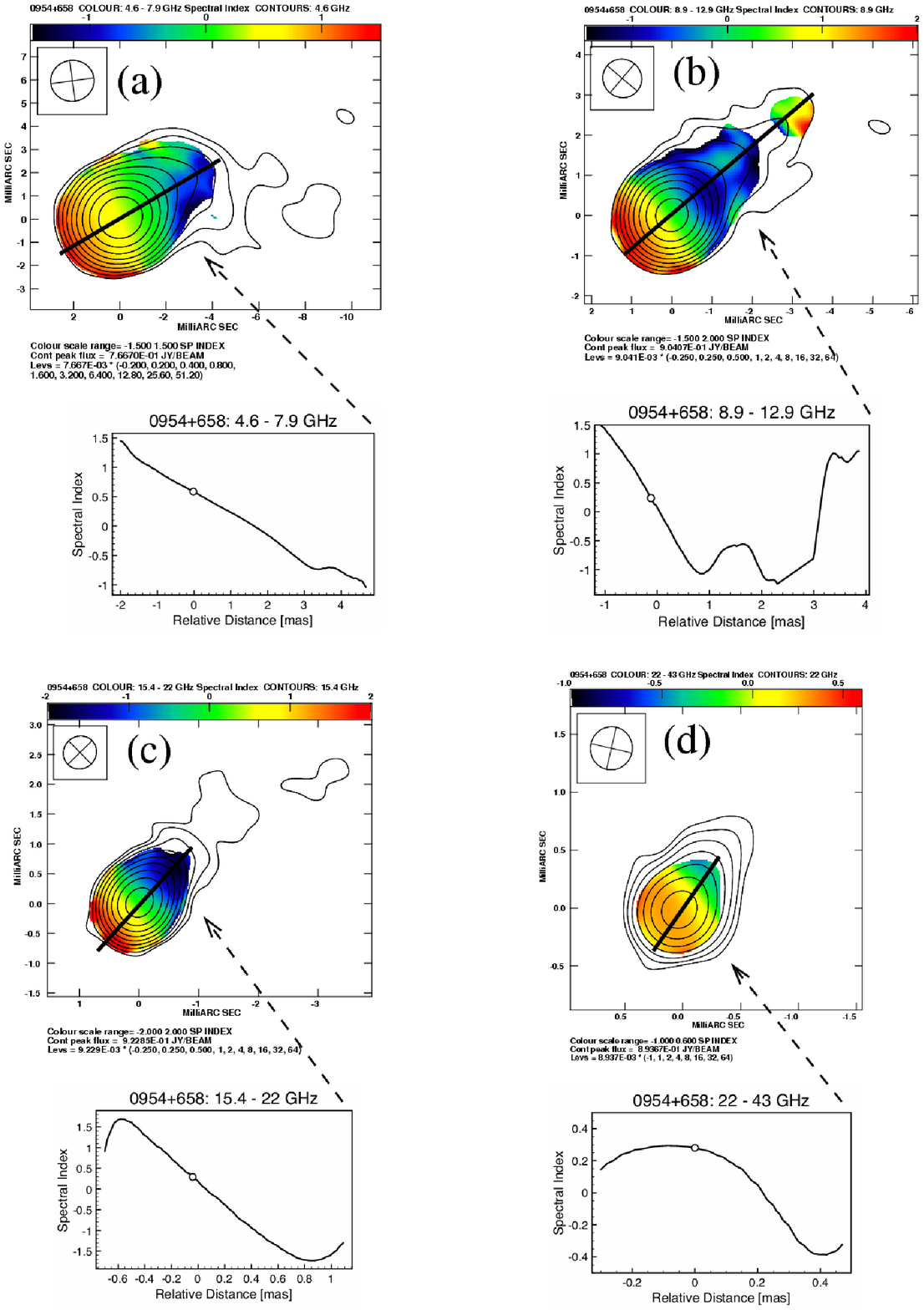}
 \caption{Spectral index maps for 0954+658 constructed from frequency pairs, with plots of slices along the jet region indicated by the solid line.
 (a) 4.6--7.9 GHz. Spectral index scale range: $-1.5$ to 1.5.
 Clip levels of {\emph I} maps: 1.5 mJy (4.6 GHz), 3.8 mJy (7.9 GHz).
 Beam FWHM and Position Angle: 1.87$\times$1.76 mas, $-82.4^{\circ}$ (4.6 GHz).
 Contours: 4.6 GHz {\emph I}, starting at 1.53 mJy and increasing by factors of 2.
 (b) 8.9--12.9 GHz. Spectral index scale range: $-1.5$ to 2.0.
 Clip levels of {\emph I} maps: 2.3 mJy (8.9 GHz), 2.6 mJy (12.9 GHz).
 Beam FWHM and Position Angle: 0.98$\times$0.95 mas, $49.8^{\circ}$ (8.9 GHz).
 Contours: 8.9 GHz {\emph I}, starting at 2.26 mJy and increasing by factors of 2.
 (c) 15.4--22 GHz. Spectral index scale range: $-2.0$ to 2.0.
 Clip levels of {\emph I} maps: 2.8 mJy (15.4 GHz), 7.8 mJy (22 GHz).
 Beam FWHM and Position Angle: 0.57$\times$0.54 mas, $-45.5^{\circ}$ (15.4 GHz).
 Contours: 15.4 GHz {\emph I}, starting at 2.31 mJy and increasing by factors of 2.
 (d) 22--43 GHz. Spectral index scale range: $-1.0$ to 0.6.
 Clip levels of {\emph I} maps: 8.9 mJy (22 GHz), 16.0 mJy (43 GHz).
 Beam FWHM and Position Angle: 0.37$\times$0.36 mas, $-12.9^{\circ}$ (22 GHz).
 Contours: 22 GHz {\emph I}, starting at 8.94 mJy and increasing by factors of 2.
 The arrows connect the plots of the spectral index slices to their corresponding maps. The hollow circle on the slice plot indicates the core position at the lower frequency of the pair.}
 \label{0954spix}
\end{figure*}

\begin{figure*}
\includegraphics[width=140mm]{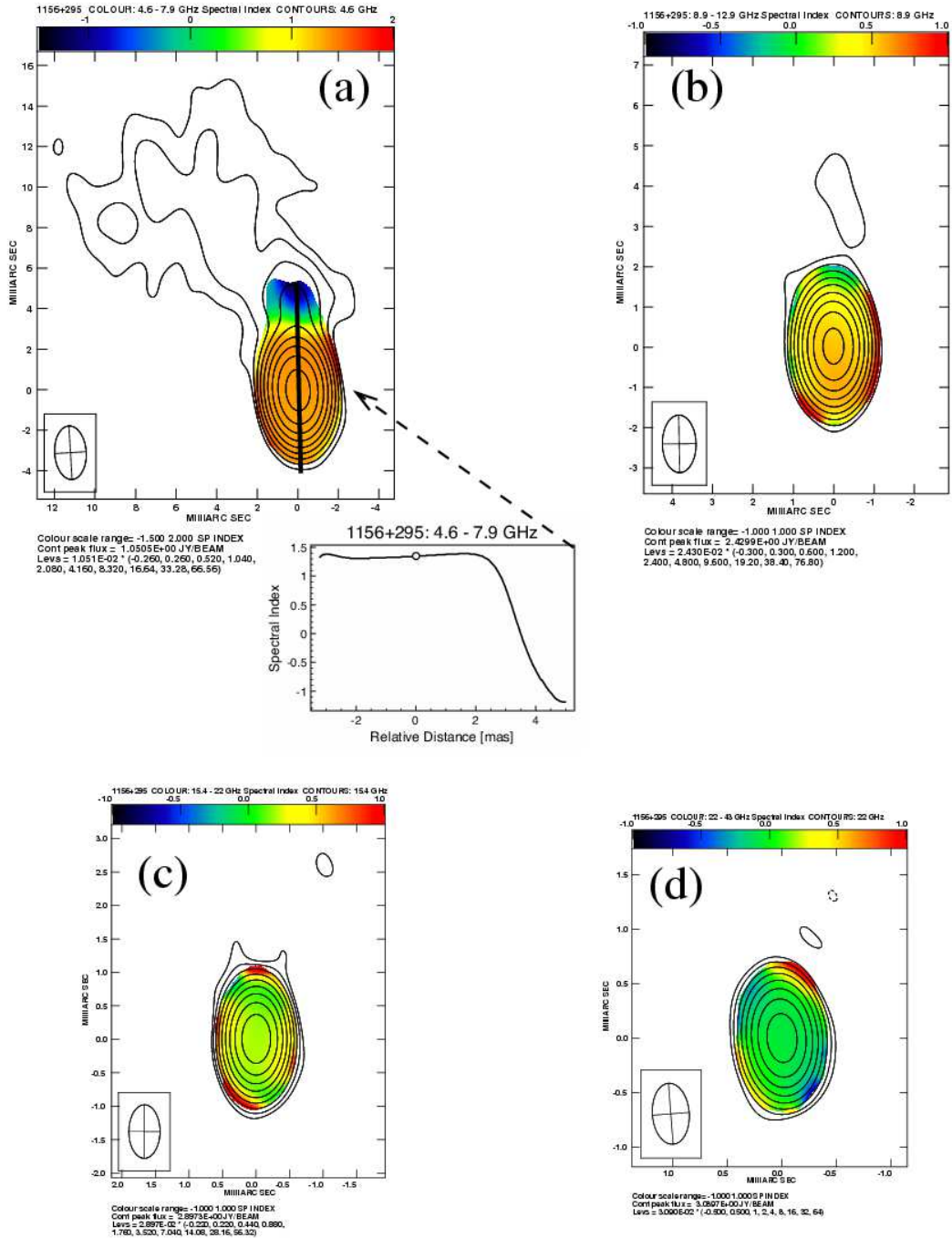}
 \caption{Spectral index maps for 1156+295 constructed from frequency pairs, with a plot of a slice along the 4.6--7.9 GHz map indicated by the solid line.
 (a) 4.6--7.9 GHz. Spectral index scale range: $-1.5$ to 2.0.
 Clip levels of {\emph I} maps: 2.7 mJy (4.6 GHz), 10.9 mJy (7.9 GHz).
 Beam FWHM and Position Angle: 2.65$\times$1.56 mas, $3.5^{\circ}$ (4.6 GHz).
 Contours: 4.6 GHz {\emph I}, starting at 2.73 mJy and increasing by factors of 2.
 (b) 8.9--12.9 GHz. Spectral index scale range: $-1.0$ to 1.0.
 Clip levels of {\emph I} maps: 7.3 mJy (8.9 GHz), 14.9 mJy (12.9 GHz).
 Beam FWHM and Position Angle: 1.43$\times$0.83 mas, $1.1^{\circ}$ (8.9 GHz).
 Contours: 8.9 GHz {\emph I}, starting at 7.29 mJy and increasing by factors of 2.
 (c) 15.4--22 GHz. Spectral index scale range: $-1.0$ to 1.0.
 Clip levels of {\emph I} maps: 5.8 mJy (15.4 GHz), 24.9 mJy (22 GHz).
 Beam FWHM and Position Angle: 0.80$\times$0.47 mas, $-0.3^{\circ}$ (15.4 GHz).
 Contours: 15.4 GHz {\emph I}, starting at 6.37 mJy and increasing by factors of 2.
 (d) 22--43 GHz. Spectral index scale range: $-1.0$ to 1.0.
 Clip levels of {\emph I} maps: 15.5 mJy (22 GHz), 44.2 mJy (43 GHz).
 Beam FWHM and Position Angle: 0.56$\times$0.34 mas, $4.4^{\circ}$ (22 GHz).
 Contours: 22 GHz {\emph I}, starting at 15.45 mJy and increasing by factors of 2.
 }
 \label{1156spix}
\end{figure*}

\newpage

\begin{figure*}
\includegraphics[width=130mm]{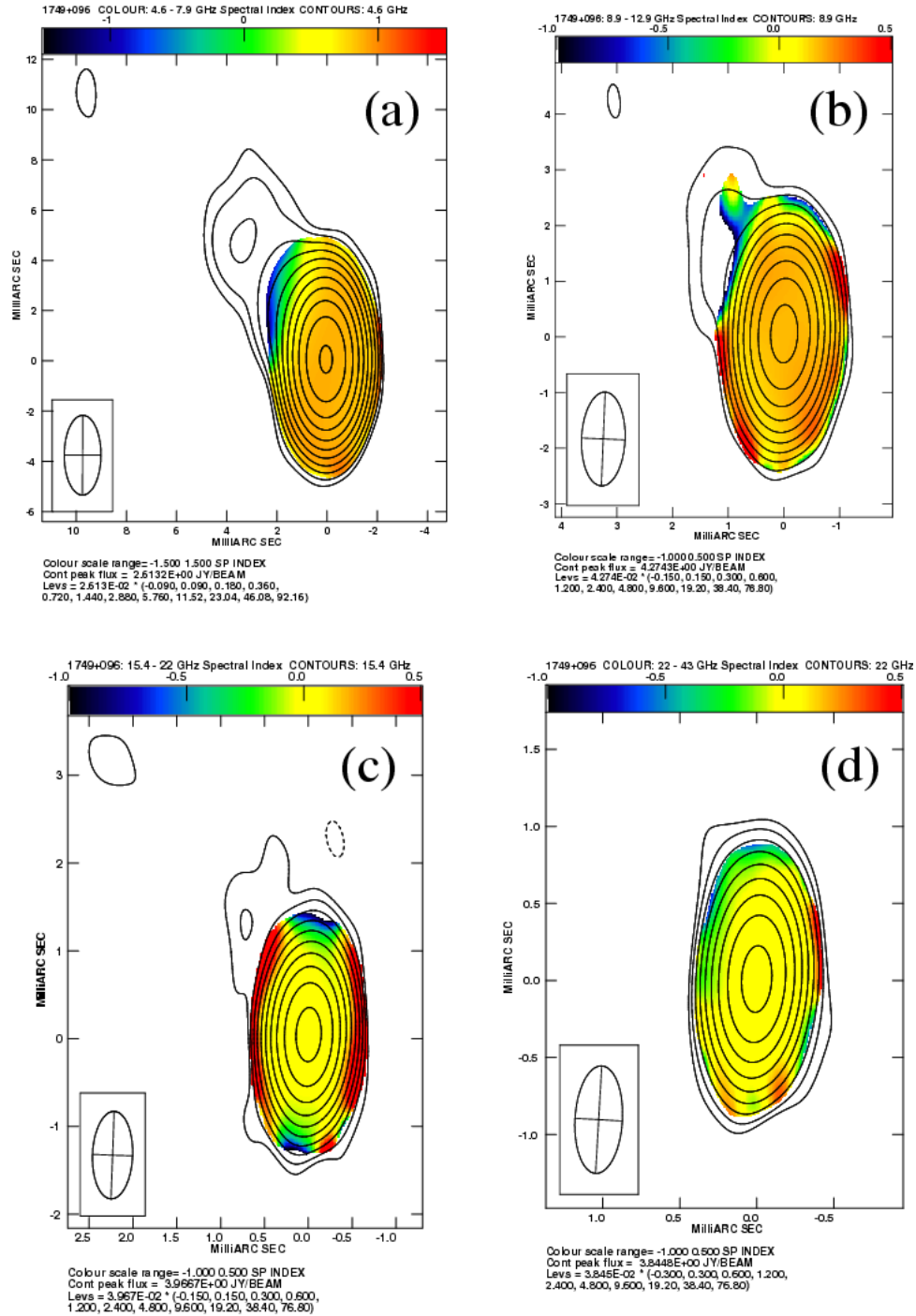}
 \caption{Spectral index maps for 1749+096 constructed from frequency pairs.
 (a) 4.6--7.9 GHz. Spectral index scale range: $-1.5$ to 1.5.
 Clip levels of {\emph I} maps: 2.4 mJy (4.6 GHz), 8.4 mJy (7.9 GHz).
 Beam FWHM and Position Angle: 3.18$\times$1.46 mas, $-0.3^{\circ}$ (4.6 GHz).
 Contours: 4.6 GHz {\emph I}, starting at 2.35 mJy and increasing by factors of 2.
 (b) 8.9--12.9 GHz. Spectral index scale range: $-1.0$ to 0.5.
 Clip levels of {\emph I} maps: 6.4 mJy (8.9 GHz), 9.1 mJy (12.9 GHz).
 Beam FWHM and Position Angle: 1.69$\times$0.78 mas, $-2.4^{\circ}$ (8.9 GHz).
 Contours: 8.9 GHz {\emph I}, starting at 6.41 mJy and increasing by factors of 2.
 (c) 15.4--22 GHz. Spectral index scale range: $-1.0$ to 0.5.
 Clip levels of {\emph I} maps: 8.3 mJy (15.4 GHz), 15.7 mJy (22 GHz).
 Beam FWHM and Position Angle: 1.00$\times$0.46 mas, $-2.2^{\circ}$ (15.4 GHz).
 Contours: 15.4 GHz {\emph I}, starting at 5.95 mJy and increasing by factors of 2.
 (d) 22--43 GHz. Spectral index scale range: $-1.0$ to 0.5.
 Clip levels of {\emph I} maps: 11.5 mJy (22 GHz), 39.3 mJy (43 GHz).
 Beam FWHM and Position Angle: 0.70$\times$0.31 mas, $-3.3^{\circ}$ (22 GHz).
 Contours: 22 GHz {\emph I}, starting at 11.54 mJy and increasing by factors of 2.
 }
 \label{1749spix}
\end{figure*}

\newpage

 

\end{document}